\documentclass[twocolumn]{aastex63}
\usepackage{graphicx}
\usepackage{natbib}
\usepackage{placeins}
\usepackage{amsmath}
\usepackage{tgcursor}
\usepackage{booktabs}
\usepackage{enumitem}
\setcitestyle{notesep={ }}
\begin{document}

\title{Astrochemistry with the Orbiting Astronomical Satellite for Investigating Stellar Systems (OASIS)}

\author[0000-0002-8716-0482]{Jennifer B. Bergner}
\correspondingauthor{Jennifer Bergner}
\email{jbergner@uchicago.edu}
\altaffiliation{NASA Sagan Fellow}
\affiliation{University of Chicago Department of the Geophysical Sciences, Chicago, IL, USA}

\author{Yancy L. Shirley}
\affiliation{Department of Astronomy and Steward Observatory, University of Arizona, Tucson, AZ, USA}

\author[0000-0001-9133-8047]{Jes K. J{\o}rgensen}
\affiliation{Niels Bohr Institute, University of Copenhagen, Copenhagen K., Denmark}

\author[0000-0003-1254-4817]{Brett McGuire}
\affiliation{Department of Chemistry, Massachusetts Institute of Technology, Cambridge, MA, USA}
\affiliation{National Radio Astronomy Observatory, Charlottesville, VA, USA}

\author[0000-0002-5828-7660]{Susanne Aalto}
\affiliation{Department of Space, Earth and Environment with Onsala Space Observatory, Chalmers University of Technology, G\"oteborg, Sweden}

\author{Carrie M. Anderson}
\affiliation{NASA Goddard Space Flight Center, Greenbelt, MD, USA}

\author{Gordon Chin}
\affiliation{NASA Goddard Space Flight Center, Greenbelt, MD, USA}

\author[0000-0002-2418-7952]{Maryvonne Gerin}
\affiliation{LERMA, Observatoire de Paris, PSL Research University, CNRS, Sorbonne Université, Paris, France}

\author{Paul Hartogh}
\affiliation{Max Planck Institute for Solar System Research, Goettingen, Germany}
 
\author{Daewook Kim}
\affiliation{Department of Astronomy and Steward Observatory, University of Arizona, Tucson, AZ, USA}
\affiliation{Wyant College of Optical Sciences, University of Arizona,Tucson, AZ, USA}
 
\author[0000-0003-1000-2547]{David Leisawitz}
\affiliation{NASA Goddard Space Flight Center, Greenbelt, MD 20771, USA}

\author{Joan Najita}
\affiliation{NSF's NOIRLab, 950 N. Cherry Avenue, Tucson, AZ, USA}

\author[0000-0002-6429-9457]{Kamber R. Schwarz}
\affiliation{Max-Planck-Institut für Astronomie, Heidelberg, Germany}
 
\author[0000-0003-0306-0028]{Alexander G.~G.~M. Tielens}
\affiliation{Astronomy Department, University of Maryland, College Park, MD, USA}
\affiliation{Leiden Observatory, University of Leiden, The Netherlands}

\author{Christopher K. Walker}
\affiliation{Department of Astronomy and Steward Observatory, University of Arizona, Tucson, AZ, USA}

\author[0000-0003-1526-7587]{David J. Wilner}
\affiliation{Center for Astrophysics \textbar\, Harvard \& Smithsonian, Cambridge, MA, USA}

\author[0000-0002-7567-4451]{Edward J. Wollack}
\affiliation{NASA Goddard Space Flight Center, Greenbelt, MD, USA}

\begin{abstract}
\noindent Chemistry along the star- and planet-formation sequence regulates how prebiotic building blocks-- carriers of the elements CHNOPS-- are incorporated into nascent planetesimals and planets.  Spectral line observations across the electromagnetic spectrum are needed to fully characterize interstellar CHNOPS chemistry, yet to date there are only limited astrochemical constraints at THz frequencies.  Here, we highlight advances to the study of CHNOPS astrochemistry that will be possible with the Orbiting Astronomical Satellite for Investigating Stellar Systems (OASIS).  OASIS is a NASA mission concept for a space-based observatory that will utilize an inflatable 14-m reflector along with a heterodyne receiver system to observe at THz frequencies with unprecedented sensitivity and angular resolution.  As part of a survey of H$_2$O and HD towards $\sim$100 protostellar and protoplanetary disk systems, OASIS will also obtain statistical constraints on the emission of complex organics from protostellar hot corinos and envelopes, as well as light hydrides including NH$_3$ and H$_2$S towards protoplanetary disks.  Line surveys of high-mass hot cores, protostellar outflow shocks, and prestellar cores will also leverage the unique capabilities of OASIS to probe high-excitation organics and small hydrides, as is needed to fully understand the chemistry of these objects.
\end{abstract}


\section{Introduction}
\label{sec:intro}

\begin{deluxetable*}{lcccc}
	\tabletypesize{\small}
	\tablecaption{OASIS receiver overview \label{tab:overview_specs}}
	\tablecolumns{5} 
	\tablewidth{\textwidth} 
	\tablehead{
        \colhead{}       & 
        \colhead{Band 1}       & 
        \colhead{Band 2} &
		\colhead{Band 3} & 
		\colhead{Band 4}          }
\startdata
Frequency range (GHz) & 455--575 & 1097--2196 & 2475--2875 & 4734--4745 \\
Beam size$^a$ ($\arcsec$) & 10.3 & 3.2 & 2.0 & 1.1 \\
Maximum IF Bandwidth (GHz) & 4/1$^b$ & 3.5 & 3.5 & 3.5 \\ 
Velocity resolution (km s$^{-1}$) & 3.1/0.03$^b$ & 1.0 & 0.6 & 0.4 \\
5$\sigma$ sensitivity$^c$ in 1h (Jy beam$^{-1}$ km s$^{-1}$) & 0.9 & 1.5 & 1.6 & 3.4 \\
\enddata
\tablenotetext{}{$^a$Diffraction-limited.  $^b$Two values are shown corresponding to medium or high spectral resolution modes.  $^c$Representative value within the band.}
\end{deluxetable*}

The elements carbon, hydrogen, nitrogen, oxygen, phosphorus, and sulfur (CHNOPS) are considered the main biogenic elements on earth, as they are found universally in all life forms.  Studying the chemistry of these elements along the star- and planet-formation sequence provides crucial insight into how they are incorporated into nascent planetesimals and planets \citep[see review by][]{Oberg2021}.  Besides regulating the bulk inventories of CHNOPS in planet-forming gas and solids, interstellar chemistry can also convert simple CHNOPS carriers into more complex organic molecules.  If this chemically complex material is incorporated into icy bodies like asteroids and comets, then it may be delivered to planetary surfaces via impact, and potentially play a role in jump-starting origins-of-life chemistry \citep[e.g.][]{Rubin2019}.  Indeed, icy bodies within the Solar system appear to be composed in part of material inherited from the early stages of star formation \citep[e.g.][]{Alexander2017, Altwegg2017, Rubin2020}, and are also implicated in the delivery of volatiles to Earth's surface \citep[e.g.][]{Alexander2012, Marty2017}.  Thus, understanding the formation and inheritance of simple and complex CHNOPS carriers along the star formation sequence is a major aim of astrochemistry.

The Orbiting Astronomical Satellite for Investigating Stellar Systems (OASIS) observatory is a NASA Medium Explorer (MIDEX) space mission concept designed to `follow the water trail' from galaxies to oceans, by covering key rotational lines of H$_2$O and HD at submillimeter wavelengths with unprecedented sensitivity and angular resolution \citep{Arenberg2021}.  With its broad frequency coverage and tunability between 455 and 4745~GHz, OASIS will also cover multitudes of spectral lines from CHNOPS carriers.  As much of our current understanding of astrochemical complexity is based on observations at millimeter wavelengths \citep{McGuire2018b}, OASIS will open a new window for studying the volatile chemistry along the star and planet formation sequence.  In particular, OASIS will transform our understanding of the astrochemistry of hydrides (e.g. NH$_\mathrm{x}$, CH$_\mathrm{x}$, OH$_\mathrm{x}$) and high-excitation lines of organic molecules.  

In this work, our aim is to highlight knowledge gaps that OASIS will uniquely fill in our understanding of organic/prebiotic astrochemistry.  For a discussion of HD and H$_2$O science applications we refer the reader to \citet{Walker2021}.  Section \ref{sec:oasis} presents a brief overview of the optical and technical capabilities of OASIS, how it compares to other state-of-the-art observing facilities, and the galactic star-forming regions it will observe.  In Section \ref{sec:astrochem} we describe how OASIS's unique observational capabilities will advance the study of CHNOPS astrochemistry along the star- and planet-formation sequence, from prestellar cores to protoplanetary disks.  Section \ref{sec:concl} presents our summary and conclusions.

\begin{figure*}
\centering
    \includegraphics[width=\linewidth]{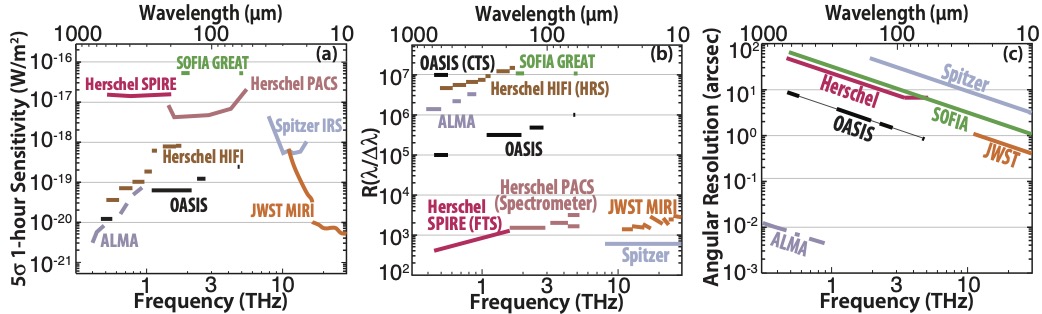}
    \caption{Comparison of (a) spectral line sensitivity, (b) spectral resolving power, and (c) angular resolution of OASIS with other state-of-the-art telescopes.}
    \label{fig:comparison_facilities}
\end{figure*}

\section{OASIS Overview}
\label{sec:oasis}

The OASIS mission concept is detailed in \citet{Walker2021}.  Here, we present a brief overview of the mission features most salient to the astrochemistry science use cases.

\subsection{Technical specifications}
OASIS is a space-based observatory, which will be in a Sun-Earth L1 halo orbit.  In the current design, OASIS utilizes an inflatable 14-m reflector, followed by aberration corrector optics achieving diffraction-limited optical performance, coupled to a state-of-the-art terahertz (THz) heterodyne receiver system \citep{Takashima2021}.  This enables high sensitivity and high spectral resolving power ($>$10$^{6}$) observations between 455 and 4745 GHz (660 and 63 $\mu$m wavelengths).  Table \ref{tab:overview_specs} summarizes the key performance characteristics of the OASIS receivers.  If selected, OASIS will launch by no later than December of 2028 at total mission cost cap of \$300 million, excluding launch costs.

\subsection{Complementarity to other facilities}
Figure \ref{fig:comparison_facilities} shows a comparison of the line sensitivity, spectral resolution, and angular resolution that OASIS will achieve, compared to other previous, existing, and future observational facilities.  The spectral line sensitivity of OASIS is comparable to that of the state-of-the-art facilities ALMA and JWST, and $\geq$10$\times$ better than \textit{Herschel} or SOFIA.  The spectral resolving power of OASIS, particularly the high-resolution chirp transform spectrometer (CTS) mode of Band 1, is comparable to that of SOFIA, ALMA, and \textit{Herschel}.  Notably, OASIS will provide 2--3 orders of magnitude higher spectral resolving power compared to JWST's NIRSPEC and MIRI instruments, allowing for detailed kinematic studies.  Lastly, OASIS will provide nearly an order of magnitude improvement in spatial resolution compared to previous and existing far-infrared telescopes (`cold' Spitzer, SOFIA, and \textit{Herschel}).  Note also that while ALMA is able to cover frequencies as high as 950 GHz, such observations require exceptional observing conditions and are not practical for extended surveys.  As a space-based facility, OASIS will readily access these wavelengths. 

OASIS also has spectral coverage overlapping with the Origins Space Telescope, a potential future facility-class mission.  As proposed, Origins is a 5.9-m cryogenic telescope with three scientific instruments operating in the wavelength range 2.8 to 588 $\mu$m \citep{Leisawitz2021}.  The Origins Survey Spectrometer (OSS) would make far-IR spectroscopic measurements with maximum spectral resolving power R$\sim$3$\times$10$^{5}$ \citep{Bradford2021}.  A fourth instrument, the Heterodyne Receiver for Origins (HERO), was also studied as a way to provide higher spectral resolving power than OSS \citep[$\sim$10$^{6}$ – 10$^{7}$; ][]{Wiedner2021}. However, HERO was not included in the baseline mission concept because heterodyne detection, limited by receiver quantum noise, does not require a very cold (4.5 K) telescope like Origins.  OASIS will accomplish Origins/HERO science in a much less expensive mission.

In summary, OASIS will outperform other far-infrared facilities (\textit{Herschel}, SOFIA), and complement near/mid-infrared (JWST) and (sub-)millimeter (ALMA) facilities as well as the future mid-/far-infrared Origins Space Telescope.  These capabilities will make OASIS a powerful instrument for astrochemical studies in star- and planet-forming regions.  In Sections \ref{subsec:prestellarcores}--\ref{subsec:disks}, we provide more detailed discussions of OASIS performance relative to other observatories in the context of specific science use cases.

\subsection{Observations of galactic star-forming regions}

As part of its aim to `follow the water trail,' OASIS will observe galaxies, nearby star-forming regions, and solar system bodies.  Here, we focus on science applications in galactic star-forming regions.  OASIS will observe various classes of objects along the star-formation sequence: pre-stellar cores, low- and high-mass protostars, and protoplanetary disks.  In particular, disk systems will be a key focus of the OASIS mission, with the goal of measuring the H$_2$O content and HD-derived disk mass towards $>$100 sources spanning young embedded disks (Class 0/I) through evolved protoplanetary disks (Class II).  Hereafter, we refer to this 100-object sample as the OASIS disk survey.

Measurements of the HD and H$_2^{18}$O lines in OASIS Bands 3 and 4 require long integrations (12 hours).  Given the large simultaneous bandwidths and independent tunability of the four OASIS Bands (Table \ref{tab:overview_specs}), broad spectral regions in Bands 1 and 2 can be scanned at the same time.  Indeed, OASIS Band 1 (455--575 GHz) can be fully covered in 12 one-hour tunings to the sensitivity shown in Figure \ref{fig:comparison_facilities}a.  For Band 2 (1100--2200 GHz), 84 GHz of the 1.1 THz band can be covered in similar one-hour tunings.  This extensive spectral coverage will contain multitudes of lines of simple and complex CHNOPS carriers.  

\begin{figure*}
\centering
    \includegraphics[width=\linewidth]{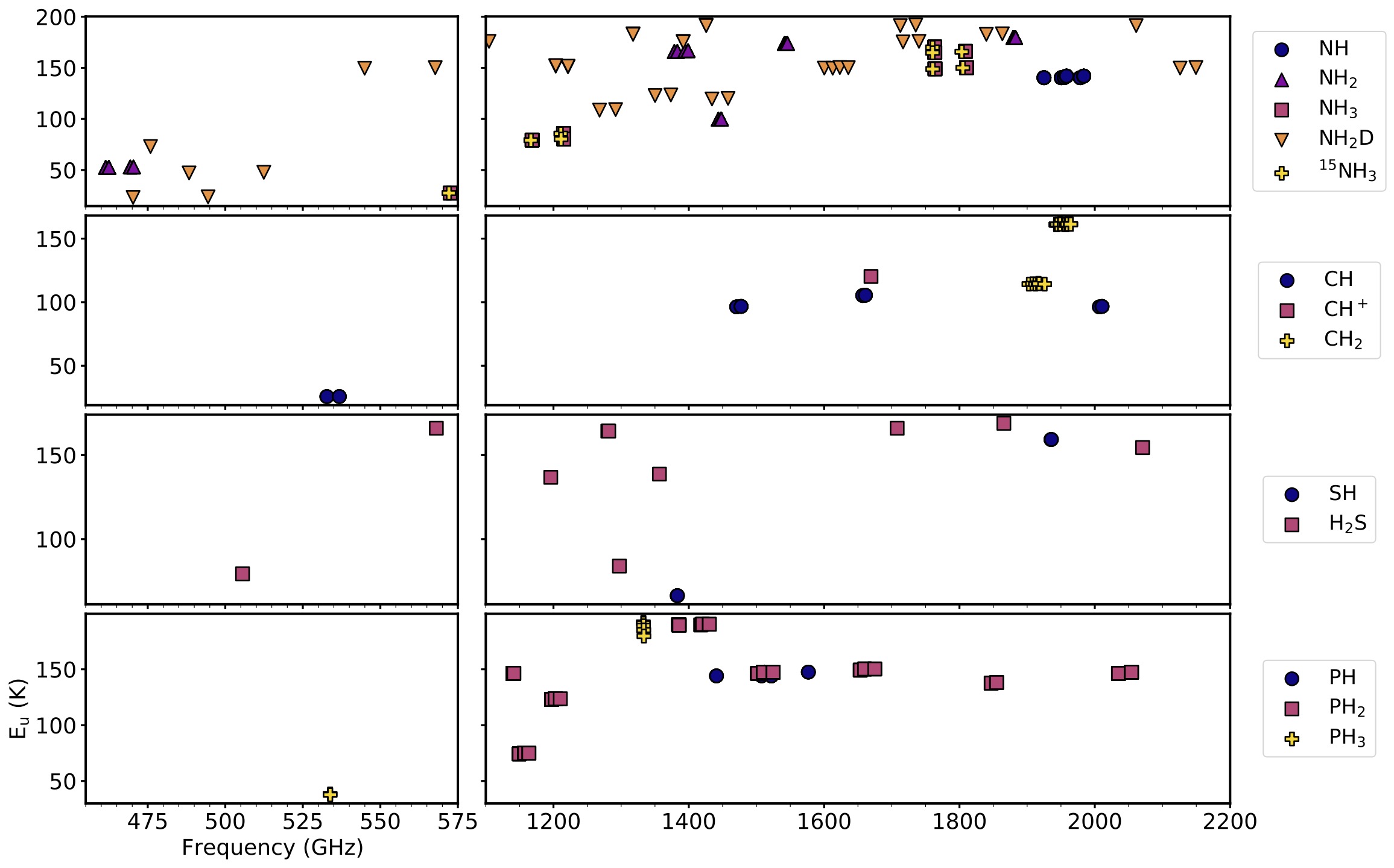}
    \caption{Summary of the frequencies and upper-state energies of selected N, C, S, and P hydrides with transitions observable by OASIS Bands 1 and 2.  Only lines with Einstein coefficients $>$10$^{-4}$ s$^{-1}$ and upper-state energies $<$200 K are included.}
    \label{fig:light_hydrides}
\end{figure*}

\section{CHNOPS astrochemistry with OASIS}
\label{sec:astrochem}
Here, we highlight impactful astrochemical contributions that OASIS will make with Band 1 and Band 2 observations of various dense star-forming regions.  We begin with objects targeted by the OASIS disk survey: Class II disks (Section \ref{subsec:disks}), and Class 0/I protostars (Section \ref{subsec:protostars}).  We next consider additional star-forming regions where line surveys with OASIS will provide novel constraints on the chemistry and physics: protostellar outflows (Section \ref{subsec:outflows}), high-mass hot cores (Section \ref{subsec:hotcores}), and prestellar cores (Section \ref{subsec:prestellarcores}). 

\subsection{Protoplanetary disks (Class II)}
\label{subsec:disks}
Given the small angular scales and cold temperatures of mature (Class II) disks, the detection of complex molecules $>$6 atoms is challenging even at lower frequencies \citep[e.g.][]{Oberg2015,Walsh2016}.  Smaller molecules are therefore essential probes of the physics and chemistry in disks, and by extension the physics and chemistry associated with planet formation.  While molecules observable at millimeter wavelengths have been extensively studied in disks, there are almost no constraints on the inventories of light hydrides in disks, many of which are observable only at sub-millimeter/FIR wavelengths.  Coverage of these lines with OASIS (Figure \ref{fig:light_hydrides}) will thus provide a novel and highly complementary avenue for exploring the volatile chemistry in disks.  Here, we highlight the light hydride science that we expect to be most impactful for studies of disk chemistry.

\subsubsection{NH$_3$ and its isotopologues}
Of the light hydrides covered by OASIS Bands 1 and 2 (Figure \ref{fig:light_hydrides}), perhaps the most exciting science will be enabled by observations of NH$_3$.  Indeed, the N budget in disks is poorly constrained given that the dominant N carrier, N$_2$, cannot be observed.  Ice spectroscopy towards low-mass protostars, the evolutionary progenitors of disks, has revealed that NH$_3$ is an important N carrier in the ice, with a relative abundance of $\sim$5\% with respect to H$_2$O compared to $<$1\% in nitriles, or XCN \citep{Oberg2011b}.  However, while nitriles are commonly detected towards disks \citep[e.g.][]{Dutrey1997, Oberg2015, Guzman2017, Bergner2019a, vanTerwisga2019}, to date NH$_3$ has been detected towards just two disks.  NH$_3$ was first detected towards the nearby TW Hya disk, via the 572 GHz transition of o-NH$_3$ observed by \textit{Herschel} \citep{Salinas2016}.  More recently, NH$_3$ was detected towards the embedded (Class I) disk GV Tau N at mid-IR wavelengths, tracing hot emission from the inner $\sim$au \citep{Najita2021}.  

\begin{figure}
\centering
    \includegraphics[width=\linewidth]{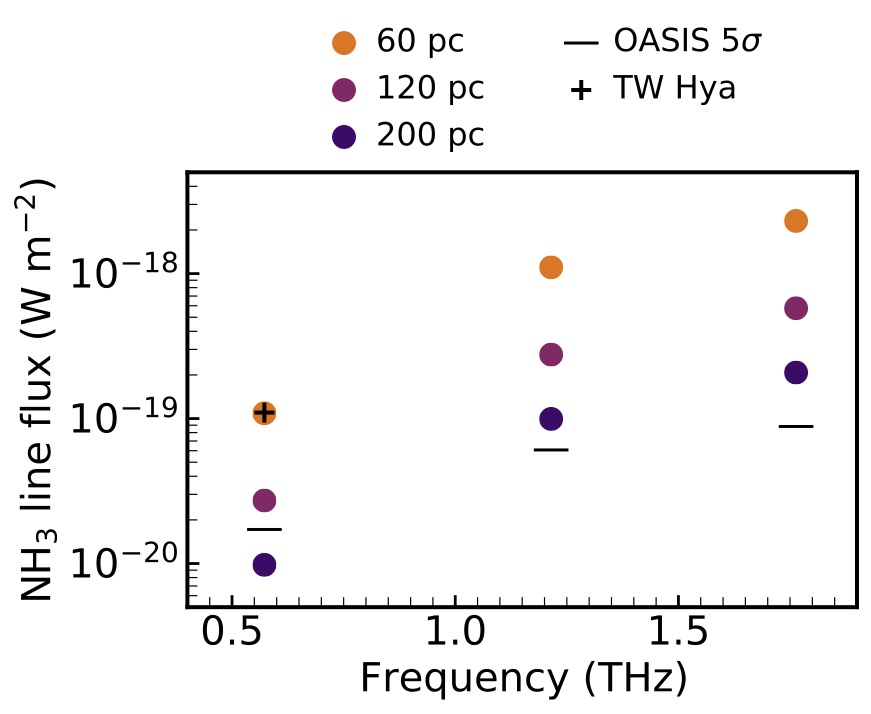}
    \caption{Simulated o-NH$_3$ line fluxes for a TW Hya-like disk at 60, 120, and 200 pc.  Horizontal bars represent the OASIS 5$\sigma$ detection threshold for each line.  The 572 GHz o-NH$_3$ flux observed by \citet{Salinas2016} is shown with a `+'.}
    \label{fig:sim_nh3_fluxes}
\end{figure}

Bands 1 and 2 of OASIS will cover multiple strong transitions tracing cool NH$_3$ (upper state energies from 27--170 K).  While the 572 GHz NH$_3$ line towards TW Hya was detected with low SNR with \textit{Herschel}, the same flux is over an order of magnitude above the OASIS 5$\sigma$ detection threshold.  To further assess the prospects of detecting NH$_3$ with OASIS, we created a toy disk model based on TW Hya.  We adopt a physical structure based on the models of \citet{Cleeves2015}, \citet{Zhang2017}, and \citet{Huang2018}.  Informed by the constraints on o-NH$_3$ from \citet{Salinas2016}, we assume a power-law abundance profile $X(r) = 8\times10^{-11}(r/100 \mathrm{au})$.  We assume a freeze-out temperature of 65K \citep[e.g.][]{Kruczkiewicz2021}, below which we attenuated the NH$_3$ abundance by two orders of magnitude.  We use RADMC-3D \citep{Dullemond2012} to simulate fluxes of the three o-NH$_3$ lines observable by OASIS, and recover a comparable flux of the 572 GHz transition compared to the \textit{Herschel} observations of TW Hya.  We emphasize that this is a toy model that reproduces the observations with a reasonable physical and chemical structure, but other distributions of NH$_3$ in TW Hya are of course possible.  

Figure \ref{fig:sim_nh3_fluxes} shows the predicted o-NH$_3$ fluxes, for three distances representative of the source targets in the OASIS disk survey.  For relatively nearby sources ($<$200 pc), we expect to detect all three o-NH$_3$ lines at high SNR.  For distant sources $\sim$200 pc, we should still detect both Band 2 transitions at moderate SNR.  While some sources may have intrinsically lower fluxes than our toy model due to e.g.~lower gas-phase NH$_3$ abundances or different temperature/density structures, it is reasonable in a sample of $\sim$100 to expect that multiple NH$_3$ lines will be detected towards numerous disks in the OASIS survey.

Excitingly, detection of multiple lines with OASIS will allow for the first excitation analysis of NH$_3$ in the outer disk.  Additionally, the high spectral resolution provided by OASIS will enable a kinematic analysis of the NH$_3$ line profiles in sources with high SNR, providing constraints on the spatial origin of emission.  Auxiliary constraints on the disk structures, provided by the OASIS observations of  CO isotopologues and HD and H$_2$O, will permit robust NH$_3$ abundance retrievals.  The NH$_3$/H$_2$O abundance ratio is of particular interest, as it can be directly compared with the ratio measured in comets to provide insights into how N is inherited by solar system bodies.

While we focused our toy model on o-NH$_3$ for simplicity, OASIS will also cover 6 transitions of p-NH$_3$ with comparable line intensities to the modeled o-NH$_3$ transitions.  While the statistical NH$_3$ ortho-to-para ratio (OPR) is 1, values $<$1 and $>$1 have been found in various interstellar environments, and are thought to correspond to formation in the gas-phase vs.~grain surface, respectively \citep[e.g.][]{Umemoto1999,Persson2012, Faure2013}.  Even for a high OPR ($\sim$2), we expect to detect multiple p-NH$_3$ lines towards nearby disks.  Thus, OASIS will measure the OPR in NH$_3$, another quantity which can be directly compared to interstellar measurements to gain insight into the formation and inheritance of NH$_3$ in planet-forming disks.  OASIS will also cover transitions of smaller nitrogen hydrides (NH and NH$_2$; Figure \ref{fig:light_hydrides}), which if detected would provide further constraints on the N budget in disks.  Moreover, probing these small N hydride radicals would elucidate the role that gas-phase radical chemistry plays in incorporating N into larger species.

Numerous transitions of NH$_3$ isotopologues (e.g.~$^{15}$NH$_3$ and NH$_2$D) are also observable by OASIS, raising the possibility of measuring isotopic fractionation levels in NH$_3$.  The case of $^{15}$NH$_3$ is particularly interesting since, to date, the $^{14}$N/$^{15}$N ratio in disks has only been measured via nitriles \citep[i.e.~CN and HCN][]{Guzman2017, HilyBlant2017}, which form through a distinct chemistry compared to NH$_3$ \citep[e.g.][]{Visser2018}.  $^{15}$NH$_3$ transitions are close in frequency, upper-state energy, and intrinsic line strength to the analogous NH$_3$ transitions (Figure \ref{fig:light_hydrides}).  Detection of these lines will likely be challenging: based on our predicted o-NH$_3$ line fluxes and given that the $^{14}$N/$^{15}$N ratio measured in disk nitriles is $\gtrsim$100, $^{15}$NH$_3$ would not be detected towards TW Hya.  Still, given the large number of disks in the OASIS survey, a detection of $^{15}$NH$_3$ is plausible and would be of high impact.

\subsubsection{Other light hydride science}
Another promising avenue for disk science with OASIS is S hydrides.  Sulfur is commonly very depleted from the gas in dense star-forming regions, though several S carriers (CS, SO, H$_2$S, and H$_2$CS) have now been detected in disks \citep{Dutrey1997, Guilloteau2013, Phuong2018, LeGal2019}.  H$_2$S was only recently detected in Class II disks: first towards GG Tau A \citep{Phuong2018}, followed by UY Aur \citep{Riviere2021}.  Towards other well-known disks, deep searches for H$_2$S have only produced upper limits \citep{Dutrey2011}.  To date only the 1$_{1,0}$--1$_{0,1}$ line (168.73 GHz) has been targeted, which is readily observable by ground-based telescopes but also intrinsically quite weak compared to the higher-frequency lines covered by OASIS.  The H$_2$S lines at 1865.6 and 1281.7 GHz appear particularly promising for detection in disks with OASIS, particularly if the emission originates in a somewhat warm environment.  For instance, assuming a comparable source-average column density (1.3$\times$10$^{12}$ cm$^{-2}$) and emitting area ($\sim$3$''$ radius) to GG Tau A, we expect to detect these lines given a rotational temperature $\gtrsim$35 K.  With warmer temperatures, closer/larger sources, or higher H$_2$S column densities, 7 additional H$_2$S lines in OASIS Bands 1 and 2 (E$_u$ 79--168 K) will also become good candidates for detection, enabling a multi-line excitation analysis.

It is important to highlight that H$_2$S is the dominant sulfur carrier in cometary comae \citep{Calmonte2016}.  The lack of constraints on H$_2$S abundances and distributions in protoplanetary disks has made it challenging to contextualize these measurements of cometary sulfur chemistry.  By providing constraints or strong upper limits on the H$_2$S inventory in a physically diverse sample of $\sim$100 disks, OASIS will greatly advance our understanding of the volatile sulfur reservoir in planet-forming disks and our own solar system.

CH$^+$ was previously detected towards one disk with \textit{Herschel}, and was found to probe warm emission from the disk surface and inner rim \citep{Thi2011}.  Prospects for detecting other light carbon hydrides (CH, CH$_2$) in disks are unclear, given that there is no precedent in the literature.  Similarly, the unsaturated nitrogen and sulfur hydrides (NH, NH$_2$, SH) have not been previously observed towards disks.  OASIS will cover numerous strong lines of each of these molecules (Figure \ref{fig:light_hydrides}), so it remains a possible avenue of discovery science.  If detected, these lines would provide yet further constraints on the volatile C/N/S chemistry in disks.  We note that phosphorus has yet to be detected in any form in a disk, and so the detection prospects for phosphorus hydrides are even less certain.

\subsection{Low-mass protostars (Class 0/I)}
\label{subsec:protostars}
\subsubsection{Hot corino emission}
A subset of low-mass protostars exhibit a rich gas-phase organic chemistry near the protostellar core, and have been termed `hot corinos' in analogy to the high-mass hot cores described in Section \ref{subsec:hotcores} \citep{Cazaux2003, Bottinelli2004}.  Because this chemical richness originates from the sublimation of icy molecules, the inventory of organic molecules in the gas should reflect, at least in part, the composition of the parent ices \citep[e.g.][]{Garrod2013}.   

\begin{figure*}
\centering
    \includegraphics[width=\linewidth]{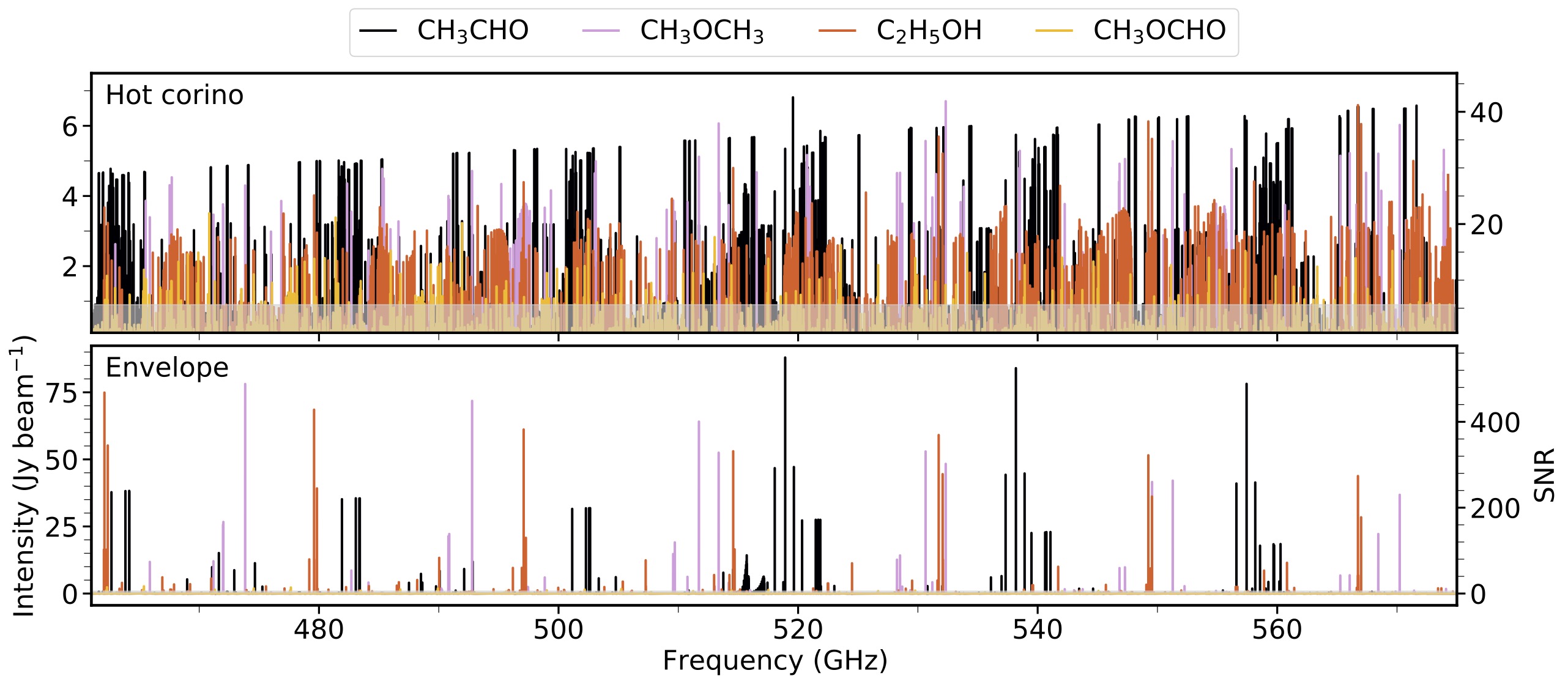}
    \caption{Simulated spectra of common O-bearing organics in OASIS Band 1.  Top and bottom panels show physical models based on a hot corino (top) and protostellar envelope (bottom).  In both panels, the 5$\sigma$ detection threshold for a one-hour integration of OASIS Band 1 is shown as a grey shaded region, assuming 1 km s$^{-1}$ line widths.}
    \label{fig:COMS_iras16293}
\end{figure*}

Within the protostellar environment, a hot corino occupies a similar spatial scale as the nascent protoplanetary disk \citep[e.g.][]{Maury2014,Maret2020}.  Constraints on the organic inventories in hot corinos therefore provide a view of the complex organic material that is present in the ices where disk formation, and ultimately planet(-esimal) formation, is taking place.  Indeed, similarities in the organic compositions of hot corinos compared to cometary ices support the idea that comets formed at least in part from icy material with a pre/protostellar origin \citep{Drozdovskaya2019}. Moreover, probing isotopic fractionation levels in these organics provides powerful constraints on their formation conditions \citep[e.g.][]{Coutens2016, Jorgensen2018}.  This is key to determining where along the star-formation sequence organic complexity is built up.

The archetypical hot corino source IRAS 16293-2422 (hereafter IRAS 16293) was observed as part of the \textit{Herschel} key project CHESS \citep{Ceccarelli2010}.  In the 555--636 GHz range, comparable to OASIS Band 1, numerous lines of small organics (e.g.~HCN, H$_2$CO, CH$_3$OH) were detected.  However, neither rarer isotopologues of CH$_3$OH nor organics larger than CH$_3$OH were detected.  With a beam size around 35$''$ at these frequencies, and a hot corino emitting size of 0.5$''$, these observations suffered a beam dilution factor of nearly 5000.  Thus, previous sub-millimeter and THz observations of hot corinos offered virtually no constraints on the organic molecule emission.

The smaller beam size of OASIS ($\sim$10$''$ at Band 1) will provide a huge improvement in this regard.  In Figure \ref{fig:COMS_iras16293} (top), we show predicted spectra of four common oxygen-bearing organics towards IRAS 16293, generated using the physical and chemical model constrained by the ALMA PILS survey \citep{Jorgensen2016}.  Although the OASIS beam will still be larger than the source size, thousands of organic molecule transitions will be detected at high SNR.  Coverage of numerous lines per molecule will permit a detailed analysis of their excitation conditions.  Moreover, with dozens of protostellar targets, the OASIS mission will deliver demographic constraints on the complex organic inventories across hot corinos.

\subsubsection{Organics in protostellar envelopes}
Not all Class 0/I protostars are hosts to hot corino emission, yet due to non-thermal desorption mechanisms, protostellar envelopes still host detectable abundances of gas-phase emission from organic molecules \citep[e.g.~][]{Oberg2011a, Bergner2017}.  Figure \ref{fig:COMS_iras16293} (bottom) shows simulated spectra assuming the same chemical model as the hot corino IRAS 16293, but adopting abundances scaled down by two orders of magnitude, and a cool (30~K) gas temperature.  For this simulation, we assume that the emission fills the OASIS beam, as is appropriate for a protostellar envelope.  Multitudes of transitions are detectable by OASIS, with even higher SNR compared to hot corinos due to the larger emitting area.  Interferometers are not well-suited to probing this diffuse envelope emission since they are mainly sensitive  to emission on smaller spatial scales, and resolve out emission on larger scales.  OASIS therefore offers a unique advantage for studying the emission of non-thermally desorbed molecules in lukewarm ($\sim$30~K) protostellar envelopes of Class 0/I sources.  These constraints will provide novel insights into how organic complexity evolves during the epoch when dust and gas of the cold envelope collapse towards the central protostar, and can help benchmark chemical models of such evolution \citep[e.g.][]{Garrod2006}.  This avenue is also highly complementary to forthcoming JWST ice observations: ice maps constructed from observations of extincted background stars will probe similar spatial scales, enabling direct comparisons between the ice and gas compositions of the envelope \citep[e.g.][]{Perotti2020, Perotti2021}.

\subsubsection{Deuterium fractionation}
\begin{figure*}
\centering
    \includegraphics[width=\linewidth]{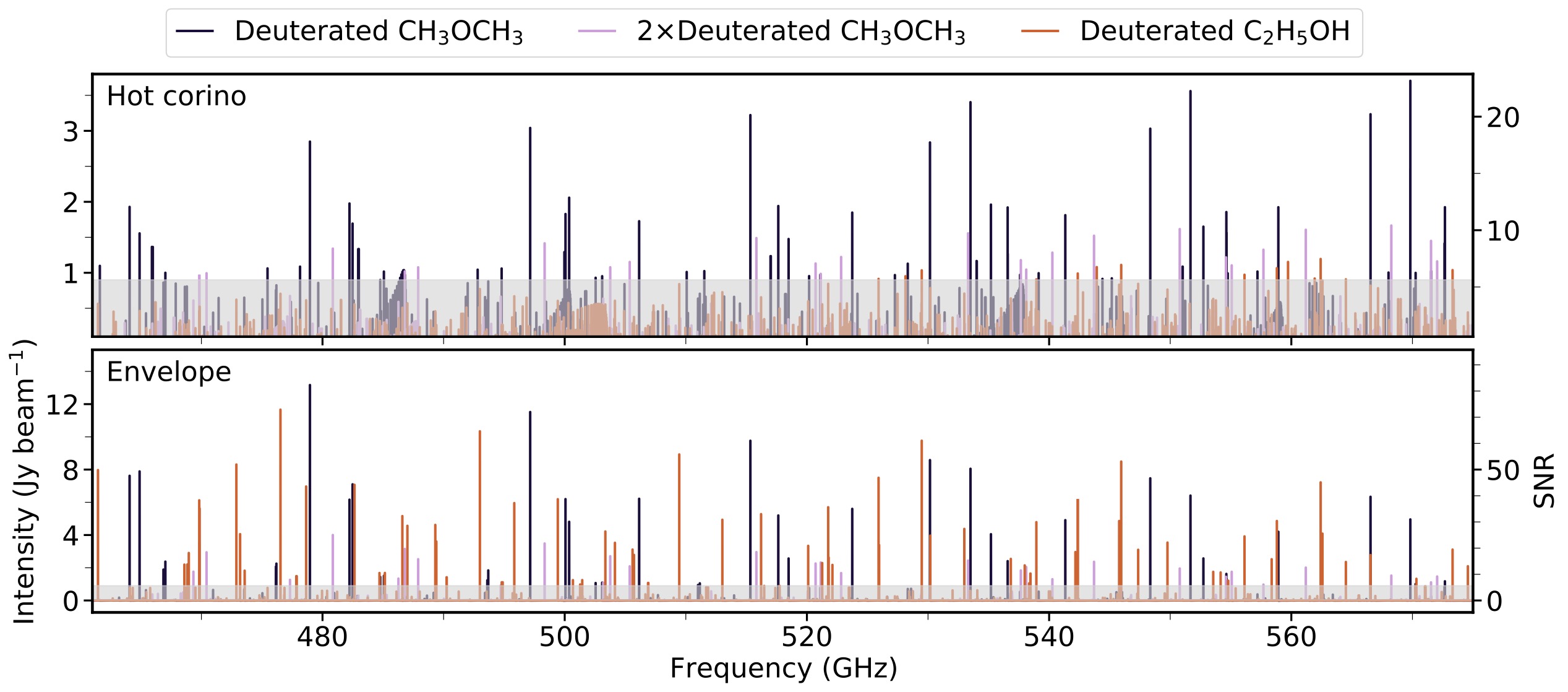}
    \caption{Simulated spectra of deuterated O-bearing organics in OASIS Band 1.  Top and bottom panels show physical models based on a hot corino and protostellar envelope, respectively.  In both panels, the 5$\sigma$ detection threshold for a one-hour integration of OASIS Band 1 is shown as a grey shaded region.}
    \label{fig:dCOMS_iras16293}
\end{figure*}

In both hot corino and non-hot corino sources, OASIS should have sufficient sensitivity to detect deuterated isotopologues of common complex organics.  Figure \ref{fig:dCOMS_iras16293} shows the simulated spectra for singly and doubly deuterated dimethyl ether (CH$_3$OCH$_3$) and singly deuterated ethanol (C$_2$H$_5$OH), assuming the same hot corino and envelope conditions as for Figure \ref{fig:COMS_iras16293}.  For clarity, different isotopomers are combined to produce a single spectrum for each substituted molecule.  While deuterated ethanol lines are only slightly above the detection threshold in the hot corino simulation, both singly and doubly deuterated dimethyl ether should be readily detectable.  The envelope scenario is even more promising, with all three molecules exhibiting numerous lines well above the detection threshold.

\subsubsection{Comparison to ALMA}
\label{subsubsec:hotcorino_alma}
It is important to consider how the performance of OASIS will compare with ALMA, the current state-of-the-art facility for studying complex chemistry in hot corinos.  Given its sub-arcsecond spatial resolution, ALMA is better beam-matched to hot corinos compared to OASIS, and is capable of detecting lower-abundance species such as even larger molecules \citep[e.g.~glycolaldehyde][]{Jorgensen2016} and rarer isotopologues \citep{Coutens2016, Jorgensen2018}.  Still, OASIS has several unique advantages.  (i) Bandwidth: In order to obtain sufficient SNR on the HD line in Band 3, OASIS will observe many sources for $\sim$12 hours.  Because Bands 1 and 2 are independently tunable, simultaneous spectral scans can cover up to 200 GHz in bandwidth assuming 1 hour per scan.  This spectral coverage is far beyond what is practically achievable with ALMA, and will provide a detailed and unbiased picture of the inventories and excitation conditions of moderately abundant complex organics.  (ii) Source statistics: The OASIS mission will observe numerous Class 0/I protostars as part of its baseline mission.  While surveys of organic chemistry in low-mass protostars are beginning to be undertaken with ALMA and NOEMA \citep[e.g.][]{Bergner2019, Belloche2020, Yang2021}, our understanding of the chemical demographics in these objects is still in its infancy.  Given also that ALMA is heavily over-subscribed, OASIS will provide much-needed statistical constraints on the complex chemistry in Class 0/I protostars. (iii) Maximum resolvable scale: As noted above, due to its larger beam size OASIS will be able to probe organic molecule emission originating in the diffuse envelope of protostars, to which ALMA observations are not sensitive.  This will provide novel constraints on chemical evolution during protostellar infall. (iv) Far-IR spectral coverage: While ALMA can access far-IR wavelengths with Bands 8--10, these observations require exceptional weather conditions.  By contrast, there are no obstacles to observing at these wavelengths with a space telescope.  Higher frequency observations generally cover lines with higher upper-state energies, and are thus well suited to studying the hot emission originating close to a protostar.  Additionally, for some molecules the spectral intensity will peak in the far-IR under hot corino conditions.  We highlight the case of E-C-cyanomethanamine, a HCN dimer implicated in prebiotic chemical schemes: at a temperature of 100 K, the strongest spectral lines occur around 900-1000 GHz \citep{Melosso2018}.  Thus, coverage at THz frequencies may be uniquely suited to detecting certain organics compared to lower-frequency observations.

\subsection{Protostellar outflows}
\label{subsec:outflows}
The early (Class 0 and Class I) stages of low-mass protostellar evolution are often accompanied by the launching of an outflow, which promotes accretion onto the protostar by carrying away angular momentum.  Encounters between the outflow and the ambient envelope material produce shocks, which can dramatically alter the local chemistry through heating and grain sputtering.  In some `chemically rich' outflows, the gas-phase abundances of molecules normally associated with the ice phase (H$_2$CO, CH$_3$OH, CH$_3$OCHO) are enhanced by orders of magnitude due to shock-induced ice sputtering \citep[e.g.][]{Garay1998, Codella1999, Requena2007, Arce2008}.  Thus, as for hot cores and hot corinos, chemically rich outflows offer a valuable window to probe the organic composition of interstellar ices.  Moreover, studies of outflow shock physics and chemistry inform our understanding of the same processes that take place on smaller, disk-forming scales within the protostellar core.

The archetypical chemically rich outflow shock, L1157-B1, was observed as part of the \textit{Herschel} CHESS survey \citep{Ceccarelli2010}.  The 555--636 GHz spectrum contained emission lines from high-excitation transitions of grain tracers like NH$_3$, H$_2$CO, and CH$_3$OH \citep{Codella2010}.  An excitation analysis of these lines revealed that they emit with temperatures $\geq$200 K, intermediate between the cold emission observed by longer-wavelength transitions and the very hot gas traced by H$_2$ emission.  Thus, observations of higher-excitation organics towards outflow shocks can help to link these different emission regimes and disentangle how the shock chemistry and physics progresses.  These insights can in turn be used to tune models of shock astrochemistry, which are needed to connect observed gas-phase abundances to the underlying grain compositions \citep[e.g.][]{Burkhardt2019}.  

\textit{Herschel} observations as part of the WISH program also provided detailed constraints on the physics and chemistry towards numerous star-forming regions including protostellar outflows \citep{vanDishoeck2011}.  While H$_2$O chemistry is not our focus, it is worth highlighting that WISH obtained extensive constraints on the abundances and origin (e.g.~ice sputtering vs.~gas-phase chemistry) of H$_2$O within outflows \citep{vanDishoeck2021}.  Additionally, light hydride lines measured through the WISH program provided powerful constraints on the UV and X-ray fields around young protostars and the physics of shock propagation \citep{Benz2016}.

OASIS will offer several improvements for observations of outflows compared to \textit{Herschel} (Figure \ref{fig:comparison_facilities}).  The improved OASIS sensitivity raises the possibility of detecting more complex species towards chemically rich shocks and characterizing their excitation in detail.  With higher angular resolution, the OASIS beam will encompass a smaller range of physical environments within the outflow, leading to less spectral blending of different shock components.  Lastly, the high-resolution mode of OASIS Band 1 has a higher spectral resolution than \textit{Herschel}, allowing different velocity components within the outflow to be clearly distinguished.  Thus, we expect significant improvements in the shock chemistry and physics that can be probed with OASIS compared to \textit{Herschel}.  While ALMA similarly offers improvements in sensitivity and spectral/spatial resolution compared to \textit{Herschel}, the higher frequencies covered by OASIS are better suited to accessing to the warm post-shocked material compared to ALMA.

It is also worth noting that chemically rich outflow shocks are, to date, the only low-mass star forming regions where phosphorus carriers have been detected \citep{Yamaguchi2011, Bergner2019b}. In shock chemistry models, PH$_3$ and smaller P hydrides are predicted to be at least as abundant as the detected P carriers PN and PO \citep{Jimenez2018}.  However, PH$_3$ has only one strong transition observable below 500 GHz, and has yet to be detected in a star-forming region.  OASIS will cover multiple strong lines of P hydrides (Figure \ref{fig:light_hydrides}), opening the door to obtaining a more complete inventory of the volatile phosphorus carriers in star-forming regions.

\subsection{High mass hot cores}
\begin{figure*}
\centering
    \includegraphics[width=\linewidth]{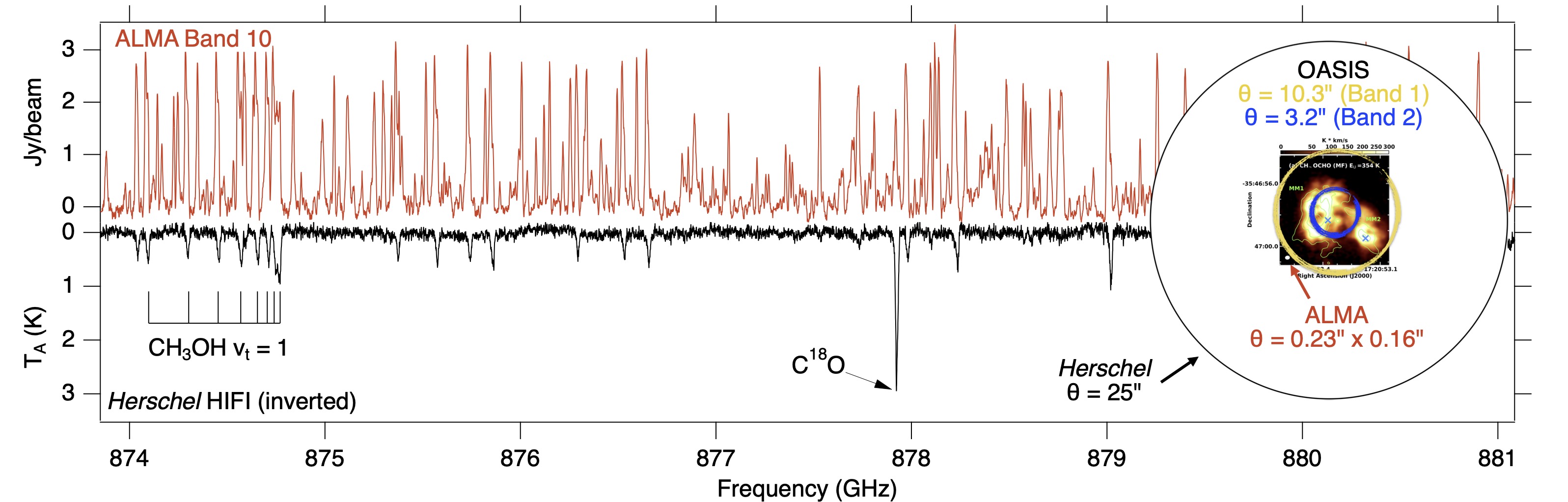}
    \caption{Spectra of the high-mass star-forming region NGC 6334I as observed by ALMA (red) compared to \textit{Herschel} (black, inverted).  Inset shows the \textit{Herschel}, ALMA, and OASIS Band 1 and Band 2 beam sizes overlaid on the source structure, as traced by methyl formate emission.  Figure adapted from \citet{McGuire2018}.}
    \label{fig:Herschel_MYSO}
\end{figure*}
\label{subsec:hotcores}
Regions of hot molecular line emission within massive star-forming regions are termed hot cores.  These regions are typically characterized by high temperatures (hundreds of K) and densities ($\sim$10$^7$ cm$^{-2}$) \citep{vanderTak2000}.  Originally identified based on the detection of hot NH$_3$ towards Orion-KL \citep{Morris1980}, hot cores were subsequently found to host an incredibly rich gas-phase organic chemistry \citep{Blake1987}.  Ice mantles are the main sites of astrochemical complex organic molecule formation, and sublimation of these ices is responsible for the wealth of chemical complexity detected in hot cores \citep[e.g.][]{Garrod2006, Herbst2009}.  The molecular line emission in hot cores provides powerful constraints on both the physical and chemical conditions in these regions \citep[see e.g.][]{Garrod2013}.  With a rich organic chemistry, high temperatures, and large emitting areas, high mass hot cores represent ideal targets to exploit the unique capabilities of OASIS.

Pioneering work in the sub-millimeter study of high-mass hot cores was performed by the \textit{Herschel} observations of EXtra-Ordinary Sources (HEXOS) guaranteed-time program.  HEXOS targeted Orion KL and Sgr B2 (N+M) with $>$1 THz in bandwidth from $\sim$480 to 1900 GHz, with spectral resolution around 1 MHz \citep{Bergin2010}.  The HEXOS program illustrates the power of wide-bandwidth, unbiased spectral scans at THz wavelengths for probing both the physics and chemistry associated with high-mass star formation.  Analyses of the sub-mm line emission and absorption generated novel constraints on the physical structure and kinematics of Orion KL and Sgr B2, including infall and stellar feedback in Sgr B2(M) \citep{Rolffs2010}; the kinematics and energetics of outflowing material in Orion KL \citep{Phillips2010}; the thermal structure and heating source in Orion KL \citep{Wang2011}; and the gas density and luminosity source within the Orion KL core \citep{Crockett2014a}.  

HEXOS provided a similar wealth of insight into the chemistry of simple molecules during high-mass star formation, including constraints on H$_2$O deuteration levels \citep{Bergin2010}; the H$_2$O abundance and ortho-to-para ratio \citep{Melnick2010}; and the CO$_2$ abundance as traced by HOCO$^+$ \citep{Neill2014}.  Additionally, many of the thousands of lines detected in the HEXOS scans correspond to complex organic molecules.  Coverage of lines with a wide range in upper-state energies allowed unprecedented constraints on the column densities and excitation conditions of dozens of organic molecules towards Orion KL and Sgr B2 \citep{Crockett2010, Crockett2014b, Neill2014}.  Intriguingly, HEXOS revealed a chemical differentiation in the complex organic inventories between Orion KL and Sgr B2(N), which could arise from evolutionary differences between the high-mass hot cores \citep{Neill2014}, and underscores the potential for such surveys to probe demographics in high-mass star formation.

While \textit{Herschel} greatly expanded our understanding of high mass hot cores, there is ample room for improvement with OASIS.  Notably, the OASIS beam will be nearly a factor of 4 smaller than the \textit{Herschel} beam at a given frequency, mitigating loss of sensitivity due to beam dilution.  For instance, in OASIS bands 1 and 2, the OASIS beam will be 10.3 and 3.2$''$ compared to 40.4 and 12.8$''$ in the corresponding \textit{Herschel} bands.  Depending on the emitting source size, this could translate to a factor of up to 16 higher line sensitivity with OASIS compared to \textit{Herschel}.  

OASIS's improved line sensitivity at THz wavelengths will open a new window into studying complex organic molecules in hot cores.  Figure \ref{fig:Herschel_MYSO}, adapted from \citet{McGuire2018}, clearly illustrates how a massive star-forming region can appear line-poor when observed by \textit{Herschel} but harbor hundreds of spectral lines when observed with a higher-sensitivity and higher-resolution observatory, in this case ALMA Band 10.  With OASIS, we similarly expect higher line densities of organic molecules compared to \textit{Herschel}.  While the increase in sensitivity will be more modest with OASIS compared to ALMA, we note that Band 9 and 10 observations with ALMA require exceptional weather conditions and also do not extend to frequencies higher than 950 GHz, whereas OASIS will provide weather-independent access to wavelengths up to 3.6 THz.

While many complex organics can be detected via lines at lower frequencies, there are several advantages to obtaining observations at far-IR wavelengths.  First, the lines covered by OASIS typically probe higher upper-state energies than millimeter-wavelength lines, which can provide a powerful lever arm for constraining excitation conditions.  This is especially important for high-mass hot cores, in which organics often have excitation temperatures of a few hundred K \citep[e.g.][]{Crockett2014b, Neill2014}.  Good constraints on organic molecule excitation temperatures are required to interpret the physical conditions of the emitting regions, as well as the chemical relationships between different classes of molecules.  Also, as noted in Section \ref{subsubsec:hotcorino_alma}, observations at higher frequencies may enable the detection of organic molecules with intensity peaks in the far-IR.

\subsection{Prestellar cores}
\label{subsec:prestellarcores}

Prestellar cores are the gravitationally bound phase of star formation immediately prior to the formation of a protostar \citep{Bergin2007, Andre2014}.
OASIS will provide a unique opportunity to probe transitions that are difficult or impossible to observe from the ground in prestellar cores.
It is during this phase that the initial conditions are set for the chemistry of the disk and subsequent planet formation.
The direct chemical inheritance from the prestellar phase to the protostellar disk has been established, for instance reflected in the D/H ratio from ALMA observations of deuterated water \citep{Jensen2021}.

\begin{figure}
\centering
    \includegraphics[scale=0.35]{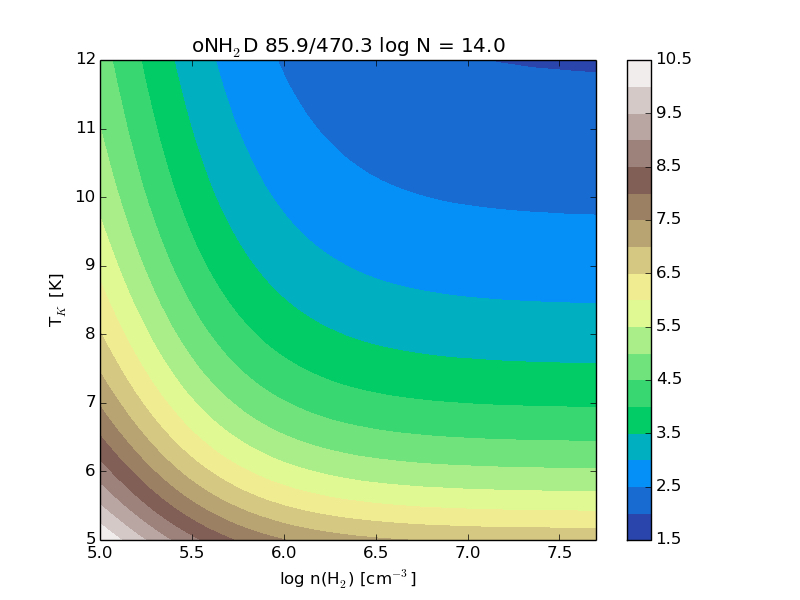}
    \caption{RADEX \citep{vanderTak2007} models of the line ratio of the 85.9 GHz to 470.3 GHz transitions of o-NH$_2$D.  Plotted line ratios vary from 1.5 to 10.5 for different values of the H$_2$ density and gas kinetic temperature.  The o-NH$_2$D column density is assumed to be $10^{14}$ cm$^{-3}$.}
    \label{fig:nh2dratio}
\end{figure}

The central region of a prestellar core is a cold ($T_K < 10$ K), dense ($n > 10^5$ cm$^{-3}$) environment that is well shielded from the surrounding interstellar radiation field (A$_{\rm{V}} > 10$ mag).
Many molecules, such as CO, deplete onto dust grains at these densities and temperatures \citep{Willacy1998, Bacmann2002} eliminating them as probes of these regions.
Deuterium fractionation of molecules via reactions with H$_2$D$^+$ become important \citep{Millar1989, Roberts2003}.
The ground state rotational transitions of para-H$_2$D$^+$ (1.370 THz) and ortho-D$_2$H$^+$ (1.477 THz) are within OASIS Band 2, however, these lines are extremely difficult to detect toward prestellar cores due to the large energy difference between rotational energy levels (i.e. $E_u/k = 65.7$ K for the upper energy level of the 1.370 THz $1_{0,1} - 0_{0,0}$ transition of para-H$_2$D$^+$).
NH$_2$D is an excellent tracer of these coldest, densest regions of prestellar cores where deuterium fractionation enhances [NH$_2$D]/[NH$_3$] by more than four orders of magnitude above the ISM D/H ratio \citep{Tine2000, Ceccarelli2014, Harju2017}.
OASIS will observe the high critical density ($n_{\rm{crit}} > 10^7$ cm$^{-3}$) transitions of NH$_2$D that probe the very centers of the cores where a disk will eventually form.
In Band 1, the $1_{1,0} - 0_{0,0}$ ground state rotation-inversion transitions of both ortho and para spin-species of NH$_2$D are observable at 470 and 494 GHz respectively (Figure \ref{fig:light_hydrides}).
These transitions may be used in combination with the 86 and 110 GHz transitions, which are observable from the ground, to constrain the physical conditions of the innermost regions.
The 86/470 line ratio of ortho-NH$_2$D is very sensitive to the temperature in the center of the core for $n(\rm{H}_2) > 10^6$ cm$^{-3}$ (Figure \ref{fig:nh2dratio}).
Observations of the 470 GHz transition are difficult from the ground because the atmospheric transmission is less than $25$\% with 1 mm of water vapor at a high altitude site and subthermal energy level populations results in a low excitation temperature and therefore weak lines.
The detectability of these lines from space were demonstrated with \textit{Herschel} observations of the prestellar cores IRAS16293E and L1544 \citep{Bacmann2012b}.
The OASIS beam will better couple to the source emission for high critical density lines in prestellar cores than the \textit{Herschel} beam permitting surveys of the densest prestellar cores in nearby molecular clouds.

Water observations of prestellar cores with OASIS will also play an important role in constraining the temperature profile in the outer part of the cores.
Most of the water in a prestellar core is found in the solid state on the icy surfaces of dust grains \citep{Bergin2002}.
However, photodesorption by UV photons can liberate some water molecules into the gas phase at abundances that are typically $< 10^{-9}$ with respect to H$_2$ \citep{vanDishoeck2021}.
There are two main sources of UV photons in prestellar cores.
The surrounding interstellar radiation field is the dominant heating component of dust grains in the core \citep{Evans2001}.
There is also a low intensity UV radiation field from excitation of H$_2$ due to collisions with electrons that come from cosmic ray ionizations of H$_2$ and He \citep{Prasad1983}.
The $1_{1,0} - 1_{0,1}$ 557 GHz ground state rotational transition of ortho-H$_2$O can be observed in absorption against the continuum of the prestellar core \citep{vanDishoeck2021}.
The line can also be seen in emission if the central density of the prestellar cores is $> 10^7$ cm$^{-3}$, although only a few prestellar cores are known that have this extreme central density \citep{Caselli2012}.
The gas phase water in the outer part of the core at low A$_{\rm{V}}$ has a photodesorption rate that depends on the strength of the interstellar radiation field (G$_0$).
A constraint on G$_0$ is needed to determine the dust temperature and the gas temperature profiles in the outer part of prestellar cores \citep{Young2004}.
Accurate temperature profiles are crucial for radiative transfer modeling of molecular emission and absorption observed toward prestellar cores.

OASIS observations of light hydride molecules will also play an important role in constraining the kinematics of prestellar cores.
The dynamical motions onto and within a prestellar core are difficult to constrain with current observational techniques.
The density profile of a hydrostatic core is very similar to the density profile of a collapsing core until very late in the collapse history \citep{Myers2005}.
Optically thick emission lines, such as HCN 1-0, can probe the internal kinematics through blue asymmetric self-absorbed emission profiles.
The creation of this line profile requires very specific kinematic and molecular excitation conditions along the line of sight that are often difficult to achieve.
Surveys searching for blue asymmetric line profiles generally find only a small fraction of prestellar cores with this signature \citep{Sohn2007, Campbell2016, Seo2019}.
An alternative strategy is to observe redshifted absorption lines from molecules that trace the outer portions of the core and compare to the velocities of emission lines that probe the inner regions of the core.
This will be possible with OASIS observations of light hydride molecules and their deuterated isotoplogues as well as with observations of H$_2$O.
Tracers that have a wide range of critical densities probe the kinematics at different depths in the core.
An example of this technique was demonstrated with \textit{Herschel} observations of the ground state rotational transitions of ND (in emission) and NH (in absorption) toward the prestellar core IRAS 16293E  \citep{Bacmann2016}.

\section{Summary and Conclusions}
\label{sec:concl}

The OASIS observatory is a NASA MIDEX space mission concept designed to observe at THz frequencies  with unprecedented sensitivity and angular resolution.  Compared to other state-of-the-art telescopes, OASIS will outperform other far-IR observatories (\textit{Herschel}, SOFIA) and complement near/mid-IR (JWST) and (sub-)millimeter (ALMA) facilities.  With its unique capacity to probe light hydrides and high-excitation organic lines, OASIS will open a new window for studying organic/prebiotic astrochemistry along the star and planet formation sequence.  Key science use cases include:

\begin{enumerate}
    \item OASIS will cover multiple strong transitions of NH$_3$ and H$_2$S, both of which are expected to be important volatiles in disks but are detected in only two sources to date.  With $\sim$100 source targets in the OASIS disk survey, OASIS will provide statistical constraints on the NH$_3$ and H$_2$S content in disks.  It will also enable, for the first time, analyses of the excitation conditions, ortho-to-para ratios, and isotopic fractionation levels of these molecules in disks.
    \item With broad spectral scans, OASIS will provide a detailed and unbiased view of the organic inventories and excitation conditions of hot corinos, lukewarm protostellar envelopes, protostellar outflows, and hot cores.  Compared to ALMA, OASIS offers important advantages in its capacity to cover wide spectral bandwidths, to obtain robust source statistics, to sample diffuse large-scale emission, and to access high-excitation lines in the far-infrared.
    \item OASIS coverage of light hydrides will provide novel constraints on the physical conditions in the dense inner regions of prestellar cores, thus probing the initial conditions of protostellar and disk physics and chemistry.  
\end{enumerate}
Importantly, these efforts directly address high-priority science questions posed by the 2020 Decadal Survey \citep{NAP2021}, including: How are potentially habitable environments formed (E-Q3a)?  And, what generates the observed chemical complexity of molecular gas (F-Q2c)?

Auxiliary laboratory efforts in THz spectroscopy will be critical to maximizing the scientific output of OASIS.  Indeed, $\sim$10\% of lines in the \textit{Herschel} spectra of Sgr B2(N) and Orion-KL were unidentified \citep{Crockett2014b, Neill2014}.  The spectral line properties of many interstellar molecules are not well-characterized above $\sim$300 GHz, and some classes of molecules (e.g.~isotopologes, vibrationally excited states, and unstable molecules) are particularly under-studied, though there are promising avenues for filling in these gaps \citep{Widicus2019}.  Along with laboratory efforts, astrochemical modeling will greatly enrich the interpretation of OASIS observations.  We expect OASIS to be particularly powerful for testing various mechanisms for ice-phase organic molecule production \citep[e.g.~][]{Garrod2006, Shingledecker2018, Jin2020, Simons2020, Carder2021}, given the detailed constraints on organic molecule inventories, abundances, and excitation conditions that will be obtained for low- and high-mass protostars and outflow shocks.

In summary, light hydrides and high-excitation organics are key players in the chemistry at various stages throughout the entire star- and planet-formation sequence.  Our current understanding of these chemical regimes is limited due to the challenges of observing at THz frequencies.  The OASIS mission will provide an unparalleled view of this astrochemistry, thus advancing our understanding of how and in what form prebiotically important material is incorporated into nascent planets and planetesimals.  

\software{
{\fontfamily{qcr}\selectfont Matplotlib} \citep{Hunter2007},
{\fontfamily{qcr}\selectfont NumPy} \citep{VanDerWalt2011},
{\fontfamily{qcr}\selectfont Radex} \citep{vanderTak2007},
{\fontfamily{qcr}\selectfont RADMC-3D}  \citep{Dullemond2012}
}

\acknowledgements 
We are grateful to the anonymous reviewers for helpful feedback on this manuscript.  J.B.B. acknowledges support from NASA through the NASA Hubble Fellowship grant \#HST-HF2-51429.001-A awarded by the Space Telescope Science Institute, which is operated by the Association of Universities for Research in Astronomy, Incorporated, under NASA contract NAS5-26555. S. A. gratefully acknowledges funding from the European Research Council (ERC) under the European Union’s Horizon 2020 research and innovation programme (grant agreement No 789410).

\FloatBarrier
\bibliography{references}

\begin{thebibliography}{}
\expandafter\ifx\csname natexlab\endcsname\relax\def\natexlab#1{#1}\fi
\providecommand{\url}[1]{\href{#1}{#1}}
\providecommand{\dodoi}[1]{doi:~\href{http://doi.org/#1}{\nolinkurl{#1}}}
\providecommand{\doeprint}[1]{\href{http://ascl.net/#1}{\nolinkurl{http://ascl.net/#1}}}
\providecommand{\doarXiv}[1]{\href{https://arxiv.org/abs/#1}{\nolinkurl{https://arxiv.org/abs/#1}}}

\bibitem[{{Alexander} {et~al.}(2012){Alexander}, {Bowden}, {Fogel}, {Howard},
  {Herd}, \& {Nittler}}]{Alexander2012}
{Alexander}, C.~M.~O., {Bowden}, R., {Fogel}, M.~L., {et~al.} 2012, Science,
  337, 721, \dodoi{10.1126/science.1223474}

\bibitem[{{Alexander} {et~al.}(2017){Alexander}, {Nittler}, {Davidson}, \&
  {Ciesla}}]{Alexander2017}
{Alexander}, C. M.~O., {Nittler}, L.~R., {Davidson}, J., \& {Ciesla}, F.~J.
  2017, Meteoritics and Planetary Science, 52, 1797, \dodoi{10.1111/maps.12891}

\bibitem[{{Altwegg} {et~al.}(2017){Altwegg}, {Balsiger}, {Berthelier},
  {Bieler}, {Calmonte}, {De Keyser}, {Fiethe}, {Fuselier}, {Gasc}, {Gombosi},
  {Owen}, {Le Roy}, {Rubin}, {S{\'e}mon}, \& {Tzou}}]{Altwegg2017}
{Altwegg}, K., {Balsiger}, H., {Berthelier}, J.~J., {et~al.} 2017,
  Philosophical Transactions of the Royal Society of London Series A, 375,
  20160253, \dodoi{10.1098/rsta.2016.0253}

\bibitem[{{Andr{\'e}} {et~al.}(2014){Andr{\'e}}, {Di Francesco},
  {Ward-Thompson}, {Inutsuka}, {Pudritz}, \& {Pineda}}]{Andre2014}
{Andr{\'e}}, P., {Di Francesco}, J., {Ward-Thompson}, D., {et~al.} 2014, in
  Protostars and Planets VI, ed. H.~{Beuther}, R.~S. {Klessen}, C.~P.
  {Dullemond}, \& T.~{Henning}, 27,
  \dodoi{10.2458/azu\_uapress\_9780816531240-ch002}

\bibitem[{{Arce} {et~al.}(2008){Arce}, {Santiago-Garc{\'\i}a}, {J{\o}rgensen},
  {Tafalla}, \& {Bachiller}}]{Arce2008}
{Arce}, H.~G., {Santiago-Garc{\'\i}a}, J., {J{\o}rgensen}, J.~K., {Tafalla},
  M., \& {Bachiller}, R. 2008, \apjl, 681, L21, \dodoi{10.1086/590110}

\bibitem[{Arenberg {et~al.}(2021)Arenberg, Villareal, Yamane, Yu, Lazear,
  Pohner, Sangalis, Jackson, Morse, Tyler, Ghosh, Palisoc, Walker, Takashima,
  Kim, Sirsi, \& Chandra}]{Arenberg2021}
Arenberg, J.~W., Villareal, M.~N., Yamane, J., {et~al.} 2021, in Astronomical
  Optics: Design, Manufacture, and Test of Space and Ground Systems III, ed.
  T.~B. Hull, D.~Kim, P.~Hallibert, \& F.~Keller, Vol. 11820, International
  Society for Optics and Photonics (SPIE), 264 -- 283.
\newblock \url{https://doi.org/10.1117/12.2594681}

\bibitem[{{Bacmann} {et~al.}(2012){Bacmann}, {Caselli}, {Ceccarelli}, {Pagani},
  \& {Vastel}}]{Bacmann2012b}
{Bacmann}, A., {Caselli}, P., {Ceccarelli}, C., {Pagani}, L., \& {Vastel}, C.
  2012, in From Atoms to Pebbles: Herschel's view of Star and Planet Formation,
  15

\bibitem[{{Bacmann} {et~al.}(2002){Bacmann}, {Lefloch}, {Ceccarelli},
  {Castets}, {Steinacker}, \& {Loinard}}]{Bacmann2002}
{Bacmann}, A., {Lefloch}, B., {Ceccarelli}, C., {et~al.} 2002, \aap, 389, L6,
  \dodoi{10.1051/0004-6361:20020652}

\bibitem[{{Bacmann} {et~al.}(2016){Bacmann}, {Daniel}, {Caselli}, {Ceccarelli},
  {Lis}, {Vastel}, {Dumouchel}, {Lique}, \& {Caux}}]{Bacmann2016}
{Bacmann}, A., {Daniel}, F., {Caselli}, P., {et~al.} 2016, \aap, 587, A26,
  \dodoi{10.1051/0004-6361/201526084}

\bibitem[{{Belloche} {et~al.}(2020){Belloche}, {Maury}, {Maret}, {Anderl},
  {Bacmann}, {Andr{\'e}}, {Bontemps}, {Cabrit}, {Codella}, {Gaudel}, {Gueth},
  {Lef{\`e}vre}, {Lefloch}, {Podio}, \& {Testi}}]{Belloche2020}
{Belloche}, A., {Maury}, A.~J., {Maret}, S., {et~al.} 2020, \aap, 635, A198,
  \dodoi{10.1051/0004-6361/201937352}

\bibitem[{{Benz} {et~al.}(2016){Benz}, {Bruderer}, {van Dishoeck}, {Melchior},
  {Wampfler}, {van der Tak}, {Goicoechea}, {Indriolo}, {Kristensen}, {Lis},
  {Mottram}, {Bergin}, {Caselli}, {Herpin}, {Hogerheijde}, {Johnstone},
  {Liseau}, {Nisini}, {Tafalla}, {Visser}, \& {Wyrowski}}]{Benz2016}
{Benz}, A.~O., {Bruderer}, S., {van Dishoeck}, E.~F., {et~al.} 2016, \aap, 590,
  A105, \dodoi{10.1051/0004-6361/201525835}

\bibitem[{{Bergin} \& {Snell}(2002)}]{Bergin2002}
{Bergin}, E.~A., \& {Snell}, R.~L. 2002, \apjl, 581, L105,
  \dodoi{10.1086/346014}

\bibitem[{{Bergin} \& {Tafalla}(2007)}]{Bergin2007}
{Bergin}, E.~A., \& {Tafalla}, M. 2007, \araa, 45, 339,
  \dodoi{10.1146/annurev.astro.45.071206.100404}

\bibitem[{{Bergin} {et~al.}(2010){Bergin}, {Phillips}, {Comito}, {Crockett},
  {Lis}, {Schilke}, {Wang}, {Bell}, {Blake}, {Bumble}, {Caux}, {Cabrit},
  {Ceccarelli}, {Cernicharo}, {Daniel}, {de Graauw}, {Dubernet},
  {Emprechtinger}, {Encrenaz}, {Falgarone}, {Gerin}, {Giesen}, {Goicoechea},
  {Goldsmith}, {Gupta}, {Hartogh}, {Helmich}, {Herbst}, {Joblin}, {Johnstone},
  {Kawamura}, {Langer}, {Latter}, {Lord}, {Maret}, {Martin}, {Melnick},
  {Menten}, {Morris}, {M{\"u}ller}, {Murphy}, {Neufeld}, {Ossenkopf}, {Pagani},
  {Pearson}, {P{\'e}rault}, {Plume}, {Roelfsema}, {Qin}, {Salez}, {Schlemmer},
  {Stutzki}, {Tielens}, {Trappe}, {van der Tak}, {Vastel}, {Yorke}, {Yu}, \&
  {Zmuidzinas}}]{Bergin2010}
{Bergin}, E.~A., {Phillips}, T.~G., {Comito}, C., {et~al.} 2010, \aap, 521,
  L20, \dodoi{10.1051/0004-6361/201015071}

\bibitem[{{Bergner} {et~al.}(2019{\natexlab{a}}){Bergner},
  {Mart{\'\i}n-Dom{\'e}nech}, {{\"O}berg}, {J{\o}rgensen}, {Artur de la
  Villarmois}, \& {Brinch}}]{Bergner2019}
{Bergner}, J.~B., {Mart{\'\i}n-Dom{\'e}nech}, R., {{\"O}berg}, K.~I., {et~al.}
  2019{\natexlab{a}}, ACS Earth and Space Chemistry, 3, 1564,
  \dodoi{10.1021/acsearthspacechem.9b00059}

\bibitem[{{Bergner} {et~al.}(2019{\natexlab{b}}){Bergner}, {{\"O}berg},
  {Bergin}, {Loomis}, {Pegues}, \& {Qi}}]{Bergner2019a}
{Bergner}, J.~B., {{\"O}berg}, K.~I., {Bergin}, E.~A., {et~al.}
  2019{\natexlab{b}}, \apj, 876, 25, \dodoi{10.3847/1538-4357/ab141e}

\bibitem[{{Bergner} {et~al.}(2017){Bergner}, {{\"O}berg}, {Garrod}, \&
  {Graninger}}]{Bergner2017}
{Bergner}, J.~B., {{\"O}berg}, K.~I., {Garrod}, R.~T., \& {Graninger}, D.~M.
  2017, \apj, 841, 120, \dodoi{10.3847/1538-4357/aa72f6}

\bibitem[{{Bergner} {et~al.}(2019{\natexlab{c}}){Bergner}, {{\"O}berg},
  {Walker}, {Guzm{\'a}n}, {Rice}, \& {Bergin}}]{Bergner2019b}
{Bergner}, J.~B., {{\"O}berg}, K.~I., {Walker}, S., {et~al.}
  2019{\natexlab{c}}, \apjl, 884, L36, \dodoi{10.3847/2041-8213/ab48f9}

\bibitem[{{Blake} {et~al.}(1987){Blake}, {Sutton}, {Masson}, \&
  {Phillips}}]{Blake1987}
{Blake}, G.~A., {Sutton}, E.~C., {Masson}, C.~R., \& {Phillips}, T.~G. 1987,
  \apj, 315, 621, \dodoi{10.1086/165165}

\bibitem[{{Bottinelli} {et~al.}(2004){Bottinelli}, {Ceccarelli}, {Lefloch},
  {Williams}, {Castets}, {Caux}, {Cazaux}, {Maret}, {Parise}, \&
  {Tielens}}]{Bottinelli2004}
{Bottinelli}, S., {Ceccarelli}, C., {Lefloch}, B., {et~al.} 2004, \apj, 615,
  354, \dodoi{10.1086/423952}

\bibitem[{{Bradford} {et~al.}(2021){Bradford}, {Cameron}, {Moore},
  {Hailey-Dunsheath}, {Amatucci}, {Bradley}, {Corsetti}, {Leisawitz},
  {DiPirro}, {Tuttle}, {Brown}, {McBirney}, {Pope}, {Armus}, {Meixner}, \&
  {Pontoppidan}}]{Bradford2021}
{Bradford}, C.~M., {Cameron}, B., {Moore}, B., {et~al.} 2021, Journal of
  Astronomical Telescopes, Instruments, and Systems, 7, 011017,
  \dodoi{10.1117/1.JATIS.7.1.011017}

\bibitem[{{Burkhardt} {et~al.}(2019){Burkhardt}, {Shingledecker}, {Le Gal},
  {McGuire}, {Remijan}, \& {Herbst}}]{Burkhardt2019}
{Burkhardt}, A.~M., {Shingledecker}, C.~N., {Le Gal}, R., {et~al.} 2019, \apj,
  881, 32, \dodoi{10.3847/1538-4357/ab2be8}

\bibitem[{{Calmonte} {et~al.}(2016){Calmonte}, {Altwegg}, {Balsiger},
  {Berthelier}, {Bieler}, {Cessateur}, {Dhooghe}, {van Dishoeck}, {Fiethe},
  {Fuselier}, {Gasc}, {Gombosi}, {H{\"a}ssig}, {Le Roy}, {Rubin}, {S{\'e}mon},
  {Tzou}, \& {Wampfler}}]{Calmonte2016}
{Calmonte}, U., {Altwegg}, K., {Balsiger}, H., {et~al.} 2016, \mnras, 462,
  S253, \dodoi{10.1093/mnras/stw2601}

\bibitem[{{Campbell} {et~al.}(2016){Campbell}, {Friesen}, {Martin}, {Caselli},
  {Kauffmann}, \& {Pineda}}]{Campbell2016}
{Campbell}, J.~L., {Friesen}, R.~K., {Martin}, P.~G., {et~al.} 2016, \apj, 819,
  143, \dodoi{10.3847/0004-637X/819/2/143}

\bibitem[{{Carder} {et~al.}(2021){Carder}, {Ochs}, \& {Herbst}}]{Carder2021}
{Carder}, J.~T., {Ochs}, W., \& {Herbst}, E. 2021, \mnras, 508, 1526,
  \dodoi{10.1093/mnras/stab2619}

\bibitem[{{Caselli} {et~al.}(2012){Caselli}, {Keto}, {Bergin}, {Tafalla},
  {Aikawa}, {Douglas}, {Pagani}, {Y{\'\i}ld{\'\i}z}, {van der Tak}, {Walmsley},
  {Codella}, {Nisini}, {Kristensen}, \& {van Dishoeck}}]{Caselli2012}
{Caselli}, P., {Keto}, E., {Bergin}, E.~A., {et~al.} 2012, \apjl, 759, L37,
  \dodoi{10.1088/2041-8205/759/2/L37}

\bibitem[{{Cazaux} {et~al.}(2003){Cazaux}, {Tielens}, {Ceccarelli}, {Castets},
  {Wakelam}, {Caux}, {Parise}, \& {Teyssier}}]{Cazaux2003}
{Cazaux}, S., {Tielens}, A.~G.~G.~M., {Ceccarelli}, C., {et~al.} 2003, \apjl,
  593, L51, \dodoi{10.1086/378038}

\bibitem[{{Ceccarelli} {et~al.}(2014){Ceccarelli}, {Caselli},
  {Bockel{\'e}e-Morvan}, {Mousis}, {Pizzarello}, {Robert}, \&
  {Semenov}}]{Ceccarelli2014}
{Ceccarelli}, C., {Caselli}, P., {Bockel{\'e}e-Morvan}, D., {et~al.} 2014, in
  Protostars and Planets VI, ed. H.~{Beuther}, R.~S. {Klessen}, C.~P.
  {Dullemond}, \& T.~{Henning}, 859,
  \dodoi{10.2458/azu\_uapress\_9780816531240-ch037}

\bibitem[{{Ceccarelli} {et~al.}(2010){Ceccarelli}, {Bacmann}, {Boogert},
  {Caux}, {Dominik}, {Lefloch}, {Lis}, {Schilke}, {van der Tak}, {Caselli},
  {Cernicharo}, {Codella}, {Comito}, {Fuente}, {Baudry}, {Bell}, {Benedettini},
  {Bergin}, {Blake}, {Bottinelli}, {Cabrit}, {Castets}, {Coutens}, {Crimier},
  {Demyk}, {Encrenaz}, {Falgarone}, {Gerin}, {Goldsmith}, {Helmich},
  {Hennebelle}, {Henning}, {Herbst}, {Hily-Blant}, {Jacq}, {Kahane}, {Kama},
  {Klotz}, {Langer}, {Lord}, {Lorenzani}, {Maret}, {Melnick}, {Neufeld},
  {Nisini}, {Pacheco}, {Pagani}, {Parise}, {Pearson}, {Phillips}, {Salez},
  {Saraceno}, {Schuster}, {Tielens}, {van der Wiel}, {Vastel}, {Viti},
  {Wakelam}, {Walters}, {Wyrowski}, {Yorke}, {Liseau}, {Olberg}, {Szczerba},
  {Benz}, \& {Melchior}}]{Ceccarelli2010}
{Ceccarelli}, C., {Bacmann}, A., {Boogert}, A., {et~al.} 2010, \aap, 521, L22,
  \dodoi{10.1051/0004-6361/201015081}

\bibitem[{{Cleeves} {et~al.}(2015){Cleeves}, {Bergin}, {Qi}, {Adams}, \&
  {{\"O}berg}}]{Cleeves2015}
{Cleeves}, L.~I., {Bergin}, E.~A., {Qi}, C., {Adams}, F.~C., \& {{\"O}berg},
  K.~I. 2015, \apj, 799, 204, \dodoi{10.1088/0004-637X/799/2/204}

\bibitem[{{Codella} \& {Bachiller}(1999)}]{Codella1999}
{Codella}, C., \& {Bachiller}, R. 1999, \aap, 350, 659

\bibitem[{{Codella} {et~al.}(2010){Codella}, {Lefloch}, {Ceccarelli},
  {Cernicharo}, {Caux}, {Lorenzani}, {Viti}, {Hily-Blant}, {Parise}, {Maret},
  {Nisini}, {Caselli}, {Cabrit}, {Pagani}, {Benedettini}, {Boogert}, {Gueth},
  {Melnick}, {Neufeld}, {Pacheco}, {Salez}, {Schuster}, {Bacmann}, {Baudry},
  {Bell}, {Bergin}, {Blake}, {Bottinelli}, {Castets}, {Comito}, {Coutens},
  {Crimier}, {Dominik}, {Demyk}, {Encrenaz}, {Falgarone}, {Fuente}, {Gerin},
  {Goldsmith}, {Helmich}, {Hennebelle}, {Henning}, {Herbst}, {Jacq}, {Kahane},
  {Kama}, {Klotz}, {Langer}, {Lis}, {Lord}, {Pearson}, {Phillips}, {Saraceno},
  {Schilke}, {Tielens}, {van der Tak}, {van der Wiel}, {Vastel}, {Wakelam},
  {Walters}, {Wyrowski}, {Yorke}, {Borys}, {Delorme}, {Kramer}, {Larsson},
  {Mehdi}, {Ossenkopf}, \& {Stutzki}}]{Codella2010}
{Codella}, C., {Lefloch}, B., {Ceccarelli}, C., {et~al.} 2010, \aap, 518, L112,
  \dodoi{10.1051/0004-6361/201014582}

\bibitem[{{Coutens} {et~al.}(2016){Coutens}, {J{\o}rgensen}, {van der Wiel},
  {M{\"u}ller}, {Lykke}, {Bjerkeli}, {Bourke}, {Calcutt}, {Drozdovskaya},
  {Favre}, {Fayolle}, {Garrod}, {Jacobsen}, {Ligterink}, {{\"O}berg},
  {Persson}, {van Dishoeck}, \& {Wampfler}}]{Coutens2016}
{Coutens}, A., {J{\o}rgensen}, J.~K., {van der Wiel}, M.~H.~D., {et~al.} 2016,
  \aap, 590, L6, \dodoi{10.1051/0004-6361/201628612}

\bibitem[{{Crockett} {et~al.}(2014{\natexlab{a}}){Crockett}, {Bergin}, {Neill},
  {Black}, {Blake}, \& {Kleshcheva}}]{Crockett2014a}
{Crockett}, N.~R., {Bergin}, E.~A., {Neill}, J.~L., {et~al.}
  2014{\natexlab{a}}, \apj, 781, 114, \dodoi{10.1088/0004-637X/781/2/114}

\bibitem[{{Crockett} {et~al.}(2010){Crockett}, {Bergin}, {Wang}, {Lis}, {Bell},
  {Blake}, {Boogert}, {Bumble}, {Cabrit}, {Caux}, {Ceccarelli}, {Cernicharo},
  {Comito}, {Daniel}, {Dubernet}, {Emprechtinger}, {Encrenaz}, {Falgarone},
  {Gerin}, {Giesen}, {Goicoechea}, {Goldsmith}, {Gupta}, {G{\"u}sten},
  {Hartogh}, {Helmich}, {Herbst}, {Honingh}, {Joblin}, {Johnstone}, {Karpov},
  {Kawamura}, {Kooi}, {Krieg}, {Langer}, {Latter}, {Lord}, {Maret}, {Martin},
  {Melnick}, {Menten}, {Morris}, {M{\"u}ller}, {Murphy}, {Neufeld},
  {Ossenkopf}, {Pearson}, {P{\'e}rault}, {Phillips}, {Plume}, {Qin},
  {Roelfsema}, {Schieder}, {Schilke}, {Schlemmer}, {Stutzki}, {van der Tak},
  {Tielens}, {Trappe}, {Vastel}, {Yorke}, {Yu}, \& {Zmuidzinas}}]{Crockett2010}
{Crockett}, N.~R., {Bergin}, E.~A., {Wang}, S., {et~al.} 2010, \aap, 521, L21,
  \dodoi{10.1051/0004-6361/201015116}

\bibitem[{{Crockett} {et~al.}(2014{\natexlab{b}}){Crockett}, {Bergin}, {Neill},
  {Favre}, {Schilke}, {Lis}, {Bell}, {Blake}, {Cernicharo}, {Emprechtinger},
  {Esplugues}, {Gupta}, {Kleshcheva}, {Lord}, {Marcelino}, {McGuire},
  {Pearson}, {Phillips}, {Plume}, {van der Tak}, {Tercero}, \&
  {Yu}}]{Crockett2014b}
{Crockett}, N.~R., {Bergin}, E.~A., {Neill}, J.~L., {et~al.}
  2014{\natexlab{b}}, \apj, 787, 112, \dodoi{10.1088/0004-637X/787/2/112}

\bibitem[{{Drozdovskaya} {et~al.}(2019){Drozdovskaya}, {van Dishoeck}, {Rubin},
  {J{\o}rgensen}, \& {Altwegg}}]{Drozdovskaya2019}
{Drozdovskaya}, M.~N., {van Dishoeck}, E.~F., {Rubin}, M., {J{\o}rgensen},
  J.~K., \& {Altwegg}, K. 2019, \mnras, 490, 50, \dodoi{10.1093/mnras/stz2430}

\bibitem[{{Dullemond} {et~al.}(2012){Dullemond}, {Juhasz}, {Pohl}, {Sereshti},
  {Shetty}, {Peters}, {Commercon}, \& {Flock}}]{Dullemond2012}
{Dullemond}, C.~P., {Juhasz}, A., {Pohl}, A., {et~al.} 2012, {RADMC-3D: A
  multi-purpose radiative transfer tool}.
\newblock \doeprint{1202.015}

\bibitem[{{Dutrey} {et~al.}(1997){Dutrey}, {Guilloteau}, \&
  {Guelin}}]{Dutrey1997}
{Dutrey}, A., {Guilloteau}, S., \& {Guelin}, M. 1997, \aap, 317, L55

\bibitem[{{Dutrey} {et~al.}(2011){Dutrey}, {Wakelam}, {Boehler}, {Guilloteau},
  {Hersant}, {Semenov}, {Chapillon}, {Henning}, {Pi{\'e}tu}, {Launhardt},
  {Gueth}, \& {Schreyer}}]{Dutrey2011}
{Dutrey}, A., {Wakelam}, V., {Boehler}, Y., {et~al.} 2011, \aap, 535, A104,
  \dodoi{10.1051/0004-6361/201116931}

\bibitem[{{Evans} {et~al.}(2001){Evans}, {Rawlings}, {Shirley}, \&
  {Mundy}}]{Evans2001}
{Evans}, Neal~J., I., {Rawlings}, J. M.~C., {Shirley}, Y.~L., \& {Mundy}, L.~G.
  2001, \apj, 557, 193, \dodoi{10.1086/321639}

\bibitem[{{Faure} {et~al.}(2013){Faure}, {Hily-Blant}, {Le Gal}, {Rist}, \&
  {Pineau des For{\^e}ts}}]{Faure2013}
{Faure}, A., {Hily-Blant}, P., {Le Gal}, R., {Rist}, C., \& {Pineau des
  For{\^e}ts}, G. 2013, \apjl, 770, L2, \dodoi{10.1088/2041-8205/770/1/L2}

\bibitem[{{Garay} {et~al.}(1998){Garay}, {K{\"o}hnenkamp}, {Bourke},
  {Rodr{\'\i}guez}, \& {Lehtinen}}]{Garay1998}
{Garay}, G., {K{\"o}hnenkamp}, I., {Bourke}, T.~L., {Rodr{\'\i}guez}, L.~F., \&
  {Lehtinen}, K.~K. 1998, \apj, 509, 768, \dodoi{10.1086/306534}

\bibitem[{{Garrod} \& {Herbst}(2006)}]{Garrod2006}
{Garrod}, R.~T., \& {Herbst}, E. 2006, \aap, 457, 927,
  \dodoi{10.1051/0004-6361:20065560}

\bibitem[{{Garrod} \& {Widicus Weaver}(2013)}]{Garrod2013}
{Garrod}, R.~T., \& {Widicus Weaver}, S.~L. 2013, Chemical Reviews, 113, 8939,
  \dodoi{10.1021/cr400147g}

\bibitem[{{Guilloteau} {et~al.}(2013){Guilloteau}, {Di Folco}, {Dutrey},
  {Simon}, {Grosso}, \& {Pi{\'e}tu}}]{Guilloteau2013}
{Guilloteau}, S., {Di Folco}, E., {Dutrey}, A., {et~al.} 2013, \aap, 549, A92,
  \dodoi{10.1051/0004-6361/201220298}

\bibitem[{{Guzm{\'a}n} {et~al.}(2017){Guzm{\'a}n}, {{\"O}berg}, {Huang},
  {Loomis}, \& {Qi}}]{Guzman2017}
{Guzm{\'a}n}, V.~V., {{\"O}berg}, K.~I., {Huang}, J., {Loomis}, R., \& {Qi}, C.
  2017, \apj, 836, 30, \dodoi{10.3847/1538-4357/836/1/30}

\bibitem[{{Harju} {et~al.}(2017){Harju}, {Daniel}, {Sipil{\"a}}, {Caselli},
  {Pineda}, {Friesen}, {Punanova}, {G{\"u}sten}, {Wiesenfeld}, {Myers},
  {Faure}, {Hily-Blant}, {Rist}, {Rosolowsky}, {Schlemmer}, \&
  {Shirley}}]{Harju2017}
{Harju}, J., {Daniel}, F., {Sipil{\"a}}, O., {et~al.} 2017, \aap, 600, A61,
  \dodoi{10.1051/0004-6361/201628463}

\bibitem[{{Herbst} \& {van Dishoeck}(2009)}]{Herbst2009}
{Herbst}, E., \& {van Dishoeck}, E.~F. 2009, \araa, 47, 427,
  \dodoi{10.1146/annurev-astro-082708-101654}

\bibitem[{{Hily-Blant} {et~al.}(2017){Hily-Blant}, {Magalhaes}, {Kastner},
  {Faure}, {Forveille}, \& {Qi}}]{HilyBlant2017}
{Hily-Blant}, P., {Magalhaes}, V., {Kastner}, J., {et~al.} 2017, \aap, 603, L6,
  \dodoi{10.1051/0004-6361/201730524}

\bibitem[{{Huang} {et~al.}(2018){Huang}, {Andrews}, {Cleeves}, {{\"O}berg},
  {Wilner}, {Bai}, {Birnstiel}, {Carpenter}, {Hughes}, {Isella}, {P{\'e}rez},
  {Ricci}, \& {Zhu}}]{Huang2018}
{Huang}, J., {Andrews}, S.~M., {Cleeves}, L.~I., {et~al.} 2018, \apj, 852, 122,
  \dodoi{10.3847/1538-4357/aaa1e7}

\bibitem[{{Hunter}(2007)}]{Hunter2007}
{Hunter}, J.~D. 2007, Computing in Science and Engineering, 9, 90,
  \dodoi{10.1109/MCSE.2007.55}

\bibitem[{{Jensen} {et~al.}(2021){Jensen}, {J{\o}rgensen}, {Kristensen},
  {Coutens}, {van Dishoeck}, {Furuya}, {Harsono}, \& {Persson}}]{Jensen2021}
{Jensen}, S.~S., {J{\o}rgensen}, J.~K., {Kristensen}, L.~E., {et~al.} 2021,
  \aap, 650, A172, \dodoi{10.1051/0004-6361/202140560}

\bibitem[{{Jim{\'e}nez-Serra} {et~al.}(2018){Jim{\'e}nez-Serra}, {Viti},
  {Qu{\'e}nard}, \& {Holdship}}]{Jimenez2018}
{Jim{\'e}nez-Serra}, I., {Viti}, S., {Qu{\'e}nard}, D., \& {Holdship}, J. 2018,
  \apj, 862, 128, \dodoi{10.3847/1538-4357/aacdf2}

\bibitem[{{Jin} \& {Garrod}(2020)}]{Jin2020}
{Jin}, M., \& {Garrod}, R.~T. 2020, \apjs, 249, 26,
  \dodoi{10.3847/1538-4365/ab9ec8}

\bibitem[{{J{\o}rgensen} {et~al.}(2016){J{\o}rgensen}, {van der Wiel},
  {Coutens}, {Lykke}, {M{\"u}ller}, {van Dishoeck}, {Calcutt}, {Bjerkeli},
  {Bourke}, {Drozdovskaya}, {Favre}, {Fayolle}, {Garrod}, {Jacobsen},
  {{\"O}berg}, {Persson}, \& {Wampfler}}]{Jorgensen2016}
{J{\o}rgensen}, J.~K., {van der Wiel}, M.~H.~D., {Coutens}, A., {et~al.} 2016,
  \aap, 595, A117, \dodoi{10.1051/0004-6361/201628648}

\bibitem[{{J{\o}rgensen} {et~al.}(2018){J{\o}rgensen}, {M{\"u}ller}, {Calcutt},
  {Coutens}, {Drozdovskaya}, {{\"O}berg}, {Persson}, {Taquet}, {van Dishoeck},
  \& {Wampfler}}]{Jorgensen2018}
{J{\o}rgensen}, J.~K., {M{\"u}ller}, H.~S.~P., {Calcutt}, H., {et~al.} 2018,
  \aap, 620, A170, \dodoi{10.1051/0004-6361/201731667}

\bibitem[{{Kruczkiewicz} {et~al.}(2021){Kruczkiewicz}, {Vitorino}, {Congiu},
  {Theul{\'e}}, \& {Dulieu}}]{Kruczkiewicz2021}
{Kruczkiewicz}, F., {Vitorino}, J., {Congiu}, E., {Theul{\'e}}, P., \&
  {Dulieu}, F. 2021, \aap, 652, A29, \dodoi{10.1051/0004-6361/202140579}

\bibitem[{{Le Gal} {et~al.}(2019){Le Gal}, {{\"O}berg}, {Loomis}, {Pegues}, \&
  {Bergner}}]{LeGal2019}
{Le Gal}, R., {{\"O}berg}, K.~I., {Loomis}, R.~A., {Pegues}, J., \& {Bergner},
  J.~B. 2019, \apj, 876, 72, \dodoi{10.3847/1538-4357/ab1416}

\bibitem[{{Leisawitz} {et~al.}(2021){Leisawitz}, {Amatucci}, {Allen},
  {Arenberg}, {Armus}, {Battersby}, {Bauer}, {Beaman}, {Bell}, {Beltran},
  {Benford}, {Bergin}, {Bolognese}, {Bradford}, {Bradley}, {Burgarella},
  {Carey}, {Carter}, {(Danny) Chi}, {Cooray}, {Corsetti}, {D'Asto}, {De Beck},
  {Denis}, {Derkacz}, {Dewell}, {DiPirro}, {Earle}, {East}, {Edgington},
  {Ennico}, {Fantano}, {Feller}, {Folta}, {Fortney}, {Gavares}, {Generie},
  {Gerin}, {Granger}, {Greene}, {Griffiths}, {Harpole}, {Harvey}, {Helmich},
  {Hilliard}, {Howard}, {Jacoby}, {Jamil}, {Jamison}, {Kaltenegger}, {Kataria},
  {Knight}, {Knollenberg}, {Lawrence}, {Lightsey}, {Lipscy}, {Mamajek},
  {Martins}, {Mather}, {Meixner}, {Melnick}, {Milam}, {Mooney}, {Moseley},
  {Narayanan}, {Neff}, {Nguyen}, {Nordt}, {Olson}, {Padgett}, {Petach},
  {Petro}, {Pohner}, {Pontoppidan}, {Pope}, {Ramspacker}, {Rao}, {Roellig},
  {Sakon}, {Sandin}, {Sandstrom}, {Scott}, {Seals}, {Sheth}, {Sokolsky},
  {Staguhn}, {Steeves}, {Stevenson}, {Stoneking}, {Su}, {Tajdaran}, {Tompkins},
  {Vieira}, {Webster}, {Wiedner}, {Wright}, {Wu}, \&
  {Zmuidzinas}}]{Leisawitz2021}
{Leisawitz}, D., {Amatucci}, E., {Allen}, L., {et~al.} 2021, Journal of
  Astronomical Telescopes, Instruments, and Systems, 7, 011002,
  \dodoi{10.1117/1.JATIS.7.1.011002}

\bibitem[{{Maret} {et~al.}(2020){Maret}, {Maury}, {Belloche}, {Gaudel},
  {Andr{\'e}}, {Cabrit}, {Codella}, {Lef{\'e}vre}, {Podio}, {Anderl}, {Gueth},
  \& {Hennebelle}}]{Maret2020}
{Maret}, S., {Maury}, A.~J., {Belloche}, A., {et~al.} 2020, \aap, 635, A15,
  \dodoi{10.1051/0004-6361/201936798}

\bibitem[{{Marty} {et~al.}(2017){Marty}, {Altwegg}, {Balsiger}, {Bar-Nun},
  {Bekaert}, {Berthelier}, {Bieler}, {Briois}, {Calmonte}, {Combi}, {De
  Keyser}, {Fiethe}, {Fuselier}, {Gasc}, {Gombosi}, {Hansen}, {H{\"a}ssig},
  {J{\"a}ckel}, {Kopp}, {Korth}, {Le Roy}, {Mall}, {Mousis}, {Owen},
  {R{\`e}me}, {Rubin}, {S{\'e}mon}, {Tzou}, {Waite}, \& {Wurz}}]{Marty2017}
{Marty}, B., {Altwegg}, K., {Balsiger}, H., {et~al.} 2017, Science, 356, 1069,
  \dodoi{10.1126/science.aal3496}

\bibitem[{{Maury} {et~al.}(2014){Maury}, {Belloche}, {Andr{\'e}}, {Maret},
  {Gueth}, {Codella}, {Cabrit}, {Testi}, \& {Bontemps}}]{Maury2014}
{Maury}, A.~J., {Belloche}, A., {Andr{\'e}}, P., {et~al.} 2014, \aap, 563, L2,
  \dodoi{10.1051/0004-6361/201323033}

\bibitem[{{McGuire}(2018)}]{McGuire2018b}
{McGuire}, B.~A. 2018, \apjs, 239, 17, \dodoi{10.3847/1538-4365/aae5d2}

\bibitem[{{McGuire} {et~al.}(2018){McGuire}, {Brogan}, {Hunter}, {Remijan},
  {Blake}, {Burkhardt}, {Carroll}, {van Dishoeck}, {Garrod}, {Linnartz},
  {Shingledecker}, \& {Willis}}]{McGuire2018}
{McGuire}, B.~A., {Brogan}, C.~L., {Hunter}, T.~R., {et~al.} 2018, \apjl, 863,
  L35, \dodoi{10.3847/2041-8213/aad7bb}

\bibitem[{{Melnick} {et~al.}(2010){Melnick}, {Tolls}, {Neufeld}, {Bergin},
  {Phillips}, {Wang}, {Crockett}, {Bell}, {Blake}, {Cabrit}, {Caux},
  {Ceccarelli}, {Cernicharo}, {Comito}, {Daniel}, {Dubernet}, {Emprechtinger},
  {Encrenaz}, {Falgarone}, {Gerin}, {Giesen}, {Goicoechea}, {Goldsmith},
  {Herbst}, {Joblin}, {Johnstone}, {Langer}, {Latter}, {Lis}, {Lord}, {Maret},
  {Martin}, {Menten}, {Morris}, {M{\"u}ller}, {Murphy}, {Ossenkopf}, {Pagani},
  {Pearson}, {P{\'e}rault}, {Plume}, {Qin}, {Salez}, {Schilke}, {Schlemmer},
  {Stutzki}, {Trappe}, {van der Tak}, {Vastel}, {Yorke}, {Yu}, \&
  {Zmuidzinas}}]{Melnick2010}
{Melnick}, G.~J., {Tolls}, V., {Neufeld}, D.~A., {et~al.} 2010, \aap, 521, L27,
  \dodoi{10.1051/0004-6361/201015085}

\bibitem[{{Melosso} {et~al.}(2018){Melosso}, {Melli}, {Puzzarini}, {Codella},
  {Spada}, {Dore}, {Degli Esposti}, {Lefloch}, {Bachiller}, {Ceccarelli},
  {Cernicharo}, \& {Barone}}]{Melosso2018}
{Melosso}, M., {Melli}, A., {Puzzarini}, C., {et~al.} 2018, \aap, 609, A121,
  \dodoi{10.1051/0004-6361/201731972}

\bibitem[{{Millar} {et~al.}(1989){Millar}, {Bennett}, \& {Herbst}}]{Millar1989}
{Millar}, T.~J., {Bennett}, A., \& {Herbst}, E. 1989, \apj, 340, 906,
  \dodoi{10.1086/167444}

\bibitem[{{Morris} {et~al.}(1980){Morris}, {Palmer}, \&
  {Zuckerman}}]{Morris1980}
{Morris}, M., {Palmer}, P., \& {Zuckerman}, B. 1980, \apj, 237, 1,
  \dodoi{10.1086/157837}

\bibitem[{{Myers}(2005)}]{Myers2005}
{Myers}, P.~C. 2005, \apj, 623, 280, \dodoi{10.1086/428386}

\bibitem[{{Najita} {et~al.}(2021){Najita}, {Carr}, {Brittain}, {Lacy},
  {Richter}, \& {Doppmann}}]{Najita2021}
{Najita}, J.~R., {Carr}, J.~S., {Brittain}, S.~D., {et~al.} 2021, \apj, 908,
  171, \dodoi{10.3847/1538-4357/abcfc6}

\bibitem[{National Academies~of Sciences \& Medicine(2021)}]{NAP2021}
National Academies~of Sciences, E., \& Medicine. 2021, Pathways to Discovery in
  Astronomy and Astrophysics for the 2020s (Washington, DC: The National
  Academies Press), \dodoi{10.17226/26141}

\bibitem[{{Neill} {et~al.}(2014){Neill}, {Bergin}, {Lis}, {Schilke},
  {Crockett}, {Favre}, {Emprechtinger}, {Comito}, {Qin}, {Anderson},
  {Burkhardt}, {Chen}, {Harris}, {Lord}, {McGuire}, {McNeill}, {Monje},
  {Phillips}, {Steber}, {Vasyunina}, \& {Yu}}]{Neill2014}
{Neill}, J.~L., {Bergin}, E.~A., {Lis}, D.~C., {et~al.} 2014, \apj, 789, 8,
  \dodoi{10.1088/0004-637X/789/1/8}

\bibitem[{{{\"O}berg} \& {Bergin}(2021)}]{Oberg2021}
{{\"O}berg}, K.~I., \& {Bergin}, E.~A. 2021, \physrep, 893, 1,
  \dodoi{10.1016/j.physrep.2020.09.004}

\bibitem[{{{\"O}berg} {et~al.}(2011{\natexlab{a}}){{\"O}berg}, {Boogert},
  {Pontoppidan}, {van den Broek}, {van Dishoeck}, {Bottinelli}, {Blake}, \&
  {Evans}}]{Oberg2011b}
{{\"O}berg}, K.~I., {Boogert}, A.~C.~A., {Pontoppidan}, K.~M., {et~al.}
  2011{\natexlab{a}}, \apj, 740, 109, \dodoi{10.1088/0004-637X/740/2/109}

\bibitem[{{{\"O}berg} {et~al.}(2015){{\"O}berg}, {Guzm{\'a}n}, {Furuya}, {Qi},
  {Aikawa}, {Andrews}, {Loomis}, \& {Wilner}}]{Oberg2015}
{{\"O}berg}, K.~I., {Guzm{\'a}n}, V.~V., {Furuya}, K., {et~al.} 2015, \nat,
  520, 198, \dodoi{10.1038/nature14276}

\bibitem[{{{\"O}berg} {et~al.}(2011{\natexlab{b}}){{\"O}berg}, {van der Marel},
  {Kristensen}, \& {van Dishoeck}}]{Oberg2011a}
{{\"O}berg}, K.~I., {van der Marel}, N., {Kristensen}, L.~E., \& {van
  Dishoeck}, E.~F. 2011{\natexlab{b}}, \apj, 740, 14,
  \dodoi{10.1088/0004-637X/740/1/14}

\bibitem[{{Perotti} {et~al.}(2021){Perotti}, {J{\o}rgensen}, {Fraser},
  {Suutarinen}, {Kristensen}, {Rocha}, {Bjerkeli}, \&
  {Pontoppidan}}]{Perotti2021}
{Perotti}, G., {J{\o}rgensen}, J.~K., {Fraser}, H.~J., {et~al.} 2021, \aap,
  650, A168, \dodoi{10.1051/0004-6361/202039669}

\bibitem[{{Perotti} {et~al.}(2020){Perotti}, {Rocha}, {J{\o}rgensen},
  {Kristensen}, {Fraser}, \& {Pontoppidan}}]{Perotti2020}
{Perotti}, G., {Rocha}, W.~R.~M., {J{\o}rgensen}, J.~K., {et~al.} 2020, \aap,
  643, A48, \dodoi{10.1051/0004-6361/202038102}

\bibitem[{{Persson} {et~al.}(2012){Persson}, {De Luca}, {Mookerjea},
  {Olofsson}, {Black}, {Gerin}, {Herbst}, {Bell}, {Coutens}, {Godard},
  {Goicoechea}, {Hassel}, {Hily-Blant}, {Menten}, {M{\"u}ller}, {Pearson}, \&
  {Yu}}]{Persson2012}
{Persson}, C.~M., {De Luca}, M., {Mookerjea}, B., {et~al.} 2012, \aap, 543,
  A145, \dodoi{10.1051/0004-6361/201118686}

\bibitem[{{Phillips} {et~al.}(2010){Phillips}, {Bergin}, {Lis}, {Neufeld},
  {Bell}, {Wang}, {Crockett}, {Emprechtinger}, {Blake}, {Caux}, {Ceccarelli},
  {Cernicharo}, {Comito}, {Daniel}, {Dubernet}, {Encrenaz}, {Gerin}, {Giesen},
  {Goicoechea}, {Goldsmith}, {Herbst}, {Joblin}, {Johnstone}, {Langer},
  {Latter}, {Lord}, {Maret}, {Martin}, {Melnick}, {Menten}, {Morris},
  {M{\"u}ller}, {Murphy}, {Ossenkopf}, {Pearson}, {P{\'e}rault}, {Plume},
  {Qin}, {Schilke}, {Schlemmer}, {Stutzki}, {Trappe}, {van der Tak}, {Vastel},
  {Yorke}, {Yu}, {Zmuidzinas}, {Boogert}, {G{\"u}sten}, {Hartogh}, {Honingh},
  {Karpov}, {Kooi}, {Krieg}, \& {Schieder}}]{Phillips2010}
{Phillips}, T.~G., {Bergin}, E.~A., {Lis}, D.~C., {et~al.} 2010, \aap, 518,
  L109, \dodoi{10.1051/0004-6361/201014570}

\bibitem[{{Phuong} {et~al.}(2018){Phuong}, {Chapillon}, {Majumdar}, {Dutrey},
  {Guilloteau}, {Pi{\'e}tu}, {Wakelam}, {Diep}, {Tang}, {Beck}, \&
  {Bary}}]{Phuong2018}
{Phuong}, N.~T., {Chapillon}, E., {Majumdar}, L., {et~al.} 2018, \aap, 616, L5,
  \dodoi{10.1051/0004-6361/201833766}

\bibitem[{{Prasad} \& {Tarafdar}(1983)}]{Prasad1983}
{Prasad}, S.~S., \& {Tarafdar}, S.~P. 1983, \apj, 267, 603,
  \dodoi{10.1086/160896}

\bibitem[{{Requena-Torres} {et~al.}(2007){Requena-Torres}, {Marcelino},
  {Jim{\'e}nez-Serra}, {Mart{\'\i}n-Pintado}, {Mart{\'\i}n}, \&
  {Mauersberger}}]{Requena2007}
{Requena-Torres}, M.~A., {Marcelino}, N., {Jim{\'e}nez-Serra}, I., {et~al.}
  2007, \apjl, 655, L37, \dodoi{10.1086/511677}

\bibitem[{{Rivi{\`e}re-Marichalar} {et~al.}(2021){Rivi{\`e}re-Marichalar},
  {Fuente}, {Le Gal}, {Arabhavi}, {Cazaux}, {Navarro-Almaida}, {Ribas},
  {Mendigut{\'\i}a}, {Barrado}, \& {Montesinos}}]{Riviere2021}
{Rivi{\`e}re-Marichalar}, P., {Fuente}, A., {Le Gal}, R., {et~al.} 2021, \aap,
  652, A46, \dodoi{10.1051/0004-6361/202140470}

\bibitem[{{Roberts} {et~al.}(2003){Roberts}, {Herbst}, \&
  {Millar}}]{Roberts2003}
{Roberts}, H., {Herbst}, E., \& {Millar}, T.~J. 2003, \apjl, 591, L41,
  \dodoi{10.1086/376962}

\bibitem[{{Rolffs} {et~al.}(2010){Rolffs}, {Schilke}, {Comito}, {Bergin}, {van
  der Tak}, {Lis}, {Qin}, {Menten}, {G{\"u}sten}, {Bell}, {Blake}, {Caux},
  {Ceccarelli}, {Cernicharo}, {Crockett}, {Daniel}, {Dubernet},
  {Emprechtinger}, {Encrenaz}, {Gerin}, {Giesen}, {Goicoechea}, {Goldsmith},
  {Gupta}, {Herbst}, {Joblin}, {Johnstone}, {Langer}, {Latter}, {Lord},
  {Maret}, {Martin}, {Melnick}, {Morris}, {M{\"u}ller}, {Murphy}, {Ossenkopf},
  {Pearson}, {P{\'e}rault}, {Phillips}, {Plume}, {Schlemmer}, {Stutzki},
  {Trappe}, {Vastel}, {Wang}, {Yorke}, {Yu}, {Zmuidzinas}, {Diez-Gonzalez},
  {Bachiller}, {Martin-Pintado}, {Baechtold}, {Olberg}, {Nordh}, {Gill}, \&
  {Chattopadhyay}}]{Rolffs2010}
{Rolffs}, R., {Schilke}, P., {Comito}, C., {et~al.} 2010, \aap, 521, L46,
  \dodoi{10.1051/0004-6361/201015106}

\bibitem[{{Rubin} {et~al.}(2019){Rubin}, {Bekaert}, {Broadley}, {Drozdovskaya},
  \& {Wampfler}}]{Rubin2019}
{Rubin}, M., {Bekaert}, D.~V., {Broadley}, M.~W., {Drozdovskaya}, M.~N., \&
  {Wampfler}, S.~F. 2019, ACS Earth and Space Chemistry, 3, 1792,
  \dodoi{10.1021/acsearthspacechem.9b00096}

\bibitem[{{Rubin} {et~al.}(2020){Rubin}, {Engrand}, {Snodgrass}, {Weissman},
  {Altwegg}, {Busemann}, {Morbidelli}, \& {Mumma}}]{Rubin2020}
{Rubin}, M., {Engrand}, C., {Snodgrass}, C., {et~al.} 2020, \ssr, 216, 102,
  \dodoi{10.1007/s11214-020-00718-2}

\bibitem[{{Salinas} {et~al.}(2016){Salinas}, {Hogerheijde}, {Bergin},
  {Cleeves}, {Brinch}, {Blake}, {Lis}, {Melnick}, {Pani{\'c}}, {Pearson},
  {Kristensen}, {Y{\i}ld{\i}z}, \& {van Dishoeck}}]{Salinas2016}
{Salinas}, V.~N., {Hogerheijde}, M.~R., {Bergin}, E.~A., {et~al.} 2016, \aap,
  591, A122, \dodoi{10.1051/0004-6361/201628172}

\bibitem[{{Seo} {et~al.}(2019){Seo}, {Majumdar}, {Goldsmith}, {Shirley},
  {Willacy}, {Ward-Thompson}, {Friesen}, {Frayer}, {Church}, {Chung}, {Cleary},
  {Cunningham}, {Devaraj}, {Egan}, {Gaier}, {Gawande}, {Gundersen}, {Harris},
  {Kangaslahti}, {Readhead}, {Samoska}, {Sieth}, {Stennes}, {Voll}, \&
  {White}}]{Seo2019}
{Seo}, Y.~M., {Majumdar}, L., {Goldsmith}, P.~F., {et~al.} 2019, \apj, 871,
  134, \dodoi{10.3847/1538-4357/aaf887}

\bibitem[{{Shingledecker} {et~al.}(2018){Shingledecker}, {Tennis}, {Le Gal}, \&
  {Herbst}}]{Shingledecker2018}
{Shingledecker}, C.~N., {Tennis}, J., {Le Gal}, R., \& {Herbst}, E. 2018, \apj,
  861, 20, \dodoi{10.3847/1538-4357/aac5ee}

\bibitem[{{Simons} {et~al.}(2020){Simons}, {Lamberts}, \&
  {Cuppen}}]{Simons2020}
{Simons}, M.~A.~J., {Lamberts}, T., \& {Cuppen}, H.~M. 2020, \aap, 634, A52,
  \dodoi{10.1051/0004-6361/201936522}

\bibitem[{{Sohn} {et~al.}(2007){Sohn}, {Lee}, {Park}, {Lee}, {Myers}, \&
  {Lee}}]{Sohn2007}
{Sohn}, J., {Lee}, C.~W., {Park}, Y.-S., {et~al.} 2007, \apj, 664, 928,
  \dodoi{10.1086/519159}

\bibitem[{Takashima {et~al.}(2021)Takashima, Sirsi, Choi, Arenberg, Kim, \&
  Walker}]{Takashima2021}
Takashima, Y., Sirsi, S., Choi, H., {et~al.} 2021, in Astronomical Optics:
  Design, Manufacture, and Test of Space and Ground Systems III, ed. T.~B.
  Hull, D.~Kim, P.~Hallibert, \& F.~Keller, Vol. 11820, International Society
  for Optics and Photonics (SPIE), 233 -- 241.
\newblock \url{https://doi.org/10.1117/12.2594610}

\bibitem[{{Thi} {et~al.}(2011){Thi}, {M{\'e}nard}, {Meeus},
  {Martin-Za{\"\i}di}, {Woitke}, {Tatulli}, {Benisty}, {Kamp}, {Pascucci},
  {Pinte}, {Grady}, {Brittain}, {White}, {Howard}, {Sandell}, \&
  {Eiroa}}]{Thi2011}
{Thi}, W.~F., {M{\'e}nard}, F., {Meeus}, G., {et~al.} 2011, \aap, 530, L2,
  \dodoi{10.1051/0004-6361/201116678}

\bibitem[{{Tin{\'e}} {et~al.}(2000){Tin{\'e}}, {Roueff}, {Falgarone}, {Gerin},
  \& {Pineau des For{\^e}ts}}]{Tine2000}
{Tin{\'e}}, S., {Roueff}, E., {Falgarone}, E., {Gerin}, M., \& {Pineau des
  For{\^e}ts}, G. 2000, \aap, 356, 1039

\bibitem[{{Umemoto} {et~al.}(1999){Umemoto}, {Mikami}, {Yamamoto}, \&
  {Hirano}}]{Umemoto1999}
{Umemoto}, T., {Mikami}, H., {Yamamoto}, S., \& {Hirano}, N. 1999, \apjl, 525,
  L105, \dodoi{10.1086/312337}

\bibitem[{{van der Tak} {et~al.}(2007){van der Tak}, {Black}, {Sch{\"o}ier},
  {Jansen}, \& {van Dishoeck}}]{vanderTak2007}
{van der Tak}, F.~F.~S., {Black}, J.~H., {Sch{\"o}ier}, F.~L., {Jansen}, D.~J.,
  \& {van Dishoeck}, E.~F. 2007, \aap, 468, 627,
  \dodoi{10.1051/0004-6361:20066820}

\bibitem[{{van der Tak} {et~al.}(2000){van der Tak}, {van Dishoeck}, {Evans},
  \& {Blake}}]{vanderTak2000}
{van der Tak}, F. F.~S., {van Dishoeck}, E.~F., {Evans}, Neal~J., I., \&
  {Blake}, G.~A. 2000, \apj, 537, 283, \dodoi{10.1086/309011}

\bibitem[{{van der Walt} {et~al.}(2011){van der Walt}, {Colbert}, \&
  {Varoquaux}}]{VanDerWalt2011}
{van der Walt}, S., {Colbert}, S.~C., \& {Varoquaux}, G. 2011, Computing in
  Science and Engineering, 13, 22, \dodoi{10.1109/MCSE.2011.37}

\bibitem[{{van Dishoeck} {et~al.}(2011){van Dishoeck}, {Kristensen}, {Benz},
  {Bergin}, {Caselli}, {Cernicharo}, {Herpin}, {Hogerheijde}, {Johnstone},
  {Liseau}, {Nisini}, {Shipman}, {Tafalla}, {van der Tak}, {Wyrowski},
  {Aikawa}, {Bachiller}, {Baudry}, {Benedettini}, {Bjerkeli}, {Blake},
  {Bontemps}, {Braine}, {Brinch}, {Bruderer}, {Chavarr{\'\i}a}, {Codella},
  {Daniel}, {de Graauw}, {Deul}, {di Giorgio}, {Dominik}, {Doty}, {Dubernet},
  {Encrenaz}, {Feuchtgruber}, {Fich}, {Frieswijk}, {Fuente}, {Giannini},
  {Goicoechea}, {Helmich}, {Herczeg}, {Jacq}, {J{\o}rgensen}, {Karska},
  {Kaufman}, {Keto}, {Larsson}, {Lefloch}, {Lis}, {Marseille}, {McCoey},
  {Melnick}, {Neufeld}, {Olberg}, {Pagani}, {Pani{\'c}}, {Parise}, {Pearson},
  {Plume}, {Risacher}, {Salter}, {Santiago-Garc{\'\i}a}, {Saraceno},
  {St{\"a}uber}, {van Kempen}, {Visser}, {Viti}, {Walmsley}, {Wampfler}, \&
  {Y{\i}ld{\i}z}}]{vanDishoeck2011}
{van Dishoeck}, E.~F., {Kristensen}, L.~E., {Benz}, A.~O., {et~al.} 2011,
  \pasp, 123, 138, \dodoi{10.1086/658676}

\bibitem[{{van Dishoeck} {et~al.}(2021){van Dishoeck}, {Kristensen}, {Mottram},
  {Benz}, {Bergin}, {Caselli}, {Herpin}, {Hogerheijde}, {Johnstone}, {Liseau},
  {Nisini}, {Tafalla}, {van der Tak}, {Wyrowski}, {Baudry}, {Benedettini},
  {Bjerkeli}, {Blake}, {Braine}, {Bruderer}, {Cabrit}, {Cernicharo}, {Choi},
  {Coutens}, {de Graauw}, {Dominik}, {Fedele}, {Fich}, {Fuente}, {Furuya},
  {Goicoechea}, {Harsono}, {Helmich}, {Herczeg}, {Jacq}, {Karska}, {Kaufman},
  {Keto}, {Lamberts}, {Larsson}, {Leurini}, {Lis}, {Melnick}, {Neufeld},
  {Pagani}, {Persson}, {Shipman}, {Taquet}, {van Kempen}, {Walsh}, {Wampfler},
  {Y{\i}ld{\i}z}, \& {WISH Team}}]{vanDishoeck2021}
{van Dishoeck}, E.~F., {Kristensen}, L.~E., {Mottram}, J.~C., {et~al.} 2021,
  \aap, 648, A24, \dodoi{10.1051/0004-6361/202039084}

\bibitem[{{van Terwisga} {et~al.}(2019){van Terwisga}, {van Dishoeck},
  {Cazzoletti}, {Facchini}, {Trapman}, {Williams}, {Manara}, {Miotello}, {van
  der Marel}, {Ansdell}, {Hogerheijde}, {Tazzari}, \&
  {Testi}}]{vanTerwisga2019}
{van Terwisga}, S.~E., {van Dishoeck}, E.~F., {Cazzoletti}, P., {et~al.} 2019,
  \aap, 623, A150, \dodoi{10.1051/0004-6361/201834257}

\bibitem[{{Visser} {et~al.}(2018){Visser}, {Bruderer}, {Cazzoletti},
  {Facchini}, {Heays}, \& {van Dishoeck}}]{Visser2018}
{Visser}, R., {Bruderer}, S., {Cazzoletti}, P., {et~al.} 2018, \aap, 615, A75,
  \dodoi{10.1051/0004-6361/201731898}

\bibitem[{Walker {et~al.}(2021)Walker, Chin, Aalto, Anderson, Arenberg,
  Battersby, Bergin, Bergner, Biver, Bjoraker, Carr, CavaliÃ©, Beck, DiSanti,
  Hartogh, Hunt, Kim, Takashima, Kulesa, Leisawitz, Najita, Rigopoulou,
  Schwarz, Shirly, Stark, Tielens, Viti, Wilner, Wollack, \&
  Young}]{Walker2021}
Walker, C.~K., Chin, G., Aalto, S., {et~al.} 2021, in Astronomical Optics:
  Design, Manufacture, and Test of Space and Ground Systems III, ed. T.~B.
  Hull, D.~Kim, P.~Hallibert, \& F.~Keller, Vol. 11820, International Society
  for Optics and Photonics (SPIE), 181 -- 232.
\newblock \url{https://doi.org/10.1117/12.2594847}

\bibitem[{{Walsh} {et~al.}(2016){Walsh}, {Loomis}, {{\"O}berg}, {Kama}, {van 't
  Hoff}, {Millar}, {Aikawa}, {Herbst}, {Widicus Weaver}, \&
  {Nomura}}]{Walsh2016}
{Walsh}, C., {Loomis}, R.~A., {{\"O}berg}, K.~I., {et~al.} 2016, \apjl, 823,
  L10, \dodoi{10.3847/2041-8205/823/1/L10}

\bibitem[{{Wang} {et~al.}(2011){Wang}, {Bergin}, {Crockett}, {Goldsmith},
  {Lis}, {Pearson}, {Schilke}, {Bell}, {Comito}, {Blake}, {Caux}, {Ceccarelli},
  {Cernicharo}, {Daniel}, {Dubernet}, {Emprechtinger}, {Encrenaz}, {Gerin},
  {Giesen}, {Goicoechea}, {Gupta}, {Herbst}, {Joblin}, {Johnstone}, {Langer},
  {Latter}, {Lord}, {Maret}, {Martin}, {Melnick}, {Menten}, {Morris},
  {M{\"u}ller}, {Murphy}, {Neufeld}, {Ossenkopf}, {P{\'e}rault}, {Phillips},
  {Plume}, {Qin}, {Schlemmer}, {Stutzki}, {Trappe}, {van der Tak}, {Vastel},
  {Yorke}, {Yu}, \& {Zmuidzinas}}]{Wang2011}
{Wang}, S., {Bergin}, E.~A., {Crockett}, N.~R., {et~al.} 2011, \aap, 527, A95,
  \dodoi{10.1051/0004-6361/201015079}

\bibitem[{{Widicus Weaver}(2019)}]{Widicus2019}
{Widicus Weaver}, S.~L. 2019, \araa, 57, 79,
  \dodoi{10.1146/annurev-astro-091918-104438}

\bibitem[{{Wiedner} {et~al.}(2021){Wiedner}, {Aalto}, {Amatucci}, {Baryshev},
  {Battersby}, {Belitsky}, {Bergin}, {Borgo}, {Carter}, {Caux}, {Cooray},
  {Corsetti}, {De Beck}, {Delorme}, {Desmaris}, {Dipirro}, {Ellison}, {Di
  Giorgio}, {Eggens}, {Gallego}, {Gerin}, {Goldsmith}, {Goldstein}, {Helmich},
  {Herpin}, {Hills}, {Hogerheijde}, {Hunt}, {Jellema}, {Keizer}, {Krieg},
  {Kroes}, {Laporte}, {Laurens}, {Leisawitz}, {Lis}, {Martins}, {Mehdi},
  {Meixner}, {Melnick}, {Milam}, {Neufeld}, {Nguyen Tuong}, {Plume},
  {Pontoppidan}, {Quertier-Dagorn}, {Risacher}, {Staguhn}, {Tong}, {Viti}, \&
  {Wyrowski}}]{Wiedner2021}
{Wiedner}, M.~C., {Aalto}, S., {Amatucci}, E.~G., {et~al.} 2021, Journal of
  Astronomical Telescopes, Instruments, and Systems, 7, 011007,
  \dodoi{10.1117/1.JATIS.7.1.011007}

\bibitem[{{Willacy} {et~al.}(1998){Willacy}, {Langer}, \&
  {Velusamy}}]{Willacy1998}
{Willacy}, K., {Langer}, W.~D., \& {Velusamy}, T. 1998, \apjl, 507, L171,
  \dodoi{10.1086/311695}

\bibitem[{{Yamaguchi} {et~al.}(2011){Yamaguchi}, {Takano}, {Sakai}, {Sakai},
  {Liu}, {Su}, {Hirano}, {Takakuwa}, {Aikawa}, {Nomura}, \&
  {Yamamoto}}]{Yamaguchi2011}
{Yamaguchi}, T., {Takano}, S., {Sakai}, N., {et~al.} 2011, \pasj, 63, L37,
  \dodoi{10.1093/pasj/63.5.L37}

\bibitem[{{Yang} {et~al.}(2021){Yang}, {Sakai}, {Zhang}, {Murillo}, {Zhang},
  {Higuchi}, {Zeng}, {L{\'o}pez-Sepulcre}, {Yamamoto}, {Lefloch}, {Bouvier},
  {Ceccarelli}, {Hirota}, {Imai}, {Oya}, {Sakai}, \& {Watanabe}}]{Yang2021}
{Yang}, Y.-L., {Sakai}, N., {Zhang}, Y., {et~al.} 2021, \apj, 910, 20,
  \dodoi{10.3847/1538-4357/abdfd6}

\bibitem[{{Young} {et~al.}(2004){Young}, {Lee}, {Evans}, {Goldsmith}, \&
  {Doty}}]{Young2004}
{Young}, K.~E., {Lee}, J.-E., {Evans}, Neal~J., I., {Goldsmith}, P.~F., \&
  {Doty}, S.~D. 2004, \apj, 614, 252, \dodoi{10.1086/423609}

\bibitem[{{Zhang} {et~al.}(2017){Zhang}, {Bergin}, {Blake}, {Cleeves}, \&
  {Schwarz}}]{Zhang2017}
{Zhang}, K., {Bergin}, E.~A., {Blake}, G.~A., {Cleeves}, L.~I., \& {Schwarz},
  K.~R. 2017, Nature Astronomy, 1, 0130, \dodoi{10.1038/s41550-017-0130}

\end{thebibliography}
\end{document}